\documentclass[12pt, twocolumn,showpacs,preprintnumbers,aps,prl,reprint,superscriptaddress]{revtex4-1}

\usepackage{graphicx}
\usepackage{subfigure}
\usepackage{color}
\usepackage{amsmath}
\usepackage[linktocpage,colorlinks=true,linkcolor=blue,citecolor=blue,breaklinks=true]{hyperref}
\usepackage{breakcites}
\usepackage{pgfplots}
\usepackage{natbib}
\usepackage{nameref}
\usepackage{xcolor}
\usepackage{cleveref}

\newcommand{\be}{\begin{equation}}
\newcommand{\ee}{\end{equation}}

\begin{document}

\preprint{}

\title{Nonlinear interactions between an unstably stratified shear flow and a phase boundary}

\author{Srikanth Toppaladoddi}
\affiliation{All Souls College, Oxford OX1 4AL, United Kingdom}
\affiliation{Department of Physics, University of Oxford, Oxford OX1 3PU, United Kingdon}
\affiliation{Mathematical Institute, University of Oxford, OX2 6GG, United Kingdom}
\email[]{srikanth.toppaladoddi@all-souls.ox.ac.uk}

\date{\today}

\begin{abstract}
{Well-resolved numerical simulations are used to study Rayleigh-B\'enard-Poiseuille flow over an evolving phase boundary for moderate values of P\'eclet ($Pe \in \left[0, 50\right]$) and Rayleigh ($Ra \in \left[2.15 \times 10^3, 10^6\right]$) numbers. The relative effects of mean shear and buoyancy are quantified using a bulk Richardson number: $Ri_b = Ra \cdot Pr/Pe^2 \in [8.6 \times 10^{-1}, 10^4]$, where $Pr$ is the Prandtl number. For $Ri_b = \mathcal{O}(1)$, we find that the Poiseuille flow inhibits convective motions, resulting in the heat transport being only due to conduction; and, for $Ri_b \gg 1$ the flow properties and heat transport closely correspond to the purely convective case. We also find that for certain $Ra$ and $Pe$, such that $Ri_b \in \left[15,95\right]$, there is a pattern competition for convection cells with a preferred aspect ratio. Furthermore, we find travelling waves at the solid-liquid interface when $Pe \neq 0$, in qualitative agreement with other sheared convective flows in the experiments of Gilpin \emph{et al.} (\emph{J. Fluid Mech} {\bf 99}(3), pp. 619-640, 1980) and the linear stability analysis of Toppaladoddi and Wettlaufer (\emph{J. Fluid Mech.} {\bf 868}, pp. 648-665, 2019).}
\end{abstract}

\maketitle

\section{Introduction}
Fluid flows that accompany solid-liquid phase transition are ubiquitous in both the natural and engineering environments \citep{epstein1983, glicksman1986, huppert1986, worster2000, hewitt2020}. The generation of fluid motions in such situations is due to buoyancy forces generated by thermal and compositional gradients arising during solidification \citep{davis1984, dietsche1985, wettlaufer1997, worster1997, wykes2018} and/or externally imposed mean shear \citep{delves1968, delves1971, Gilpin1980, coriell1984, forth1989, feltham1999, neufeld2008, neufeld2008shear, ramudu2016, bushuk2019}. In this study, we will be concerned with the shear- and buoyancy-driven flow of a pure melt over its evolving solid phase.

Some of the first systematic investigations into the effects of a phase boundary on convective motions in a pure melt are those of \citet{davis1984} and \citet{dietsche1985}. \citet{davis1984} studied fluid motions and pattern formation in Rayleigh-B\'enard convection over a phase-changing boundary using experiments and weakly nonlinear stability theory. The primary focus of their study was on identifying different regimes in which roll, hexagonal, and mixed patterns appeared at the phase boundary. Some of the key results from their study are: (i) both the critical Rayleigh number ($Ra_c$) and the critical wavenumber ($k_c$) for the onset of convection decrease monotonically with the initial thickness of the solid phase, and asymptote to constant values for large values of the initial thickness of the solid phase; (ii) hexagonal and roll patterns on the phase boundary are observed when the initial thickness of the solid phase is large and small, respectively; and (iii) the onset of hexagonal convection at the phase boundary is accompanied by a jump in the heat flux, and thereby in the mean position of the phase boundary. The subsequent experimental study of \citet{dietsche1985} confirmed the predictions of jump in the phase-boundary position and the existence of strong hysteresis behaviour near the onset of convection. They also explored the different interfacial patterns that emerged with increasing $Ra$.

Recent studies on the coupled convection--phase-change problem have been focussed on $Ra \gg Ra_c$. \citet{esfahani2018} numerically studied the interactions between a melting isothermal solid phase and convective motions in the underlying liquid phase in two and three dimensions. A key result from their study is that the dimensionless heat flux ($\mathcal{N}$) is only weakly dependent on the Stefan number ($\mathcal{S}$), which is defined as the ratio of latent heat of fusion to the specific heat content of a material and quantifies the pace at which phase change proceeds. Using a similar configuration, \citet{favier2019} systematically explored the different transitions in the convection cell structure as the solid and liquid phases evolved. They showed that due to the presence of the phase boundary, the flow remains steady even at large $Ra$. This results in higher heat transport than in the classical Rayleigh-B\'enard convection in two dimensions, where the flow becomes unsteady at $Ra \approx 7.5 \times 10^5$ \citep{TSW2015_EPL}. \citet{purseed2020} considered a more general situation where the melting point of the solid lies between the temperatures imposed at the upper and lower boundaries, and studied the bistability close to the onset of convection which was first predicted by \citet{davis1984}. 

From the studies of thermal convection over phase boundaries it can be concluded that when the temperature of the upper boundary is less than the melting point, the phase boundary develops steady patterns -- polygons, rolls, or a mix of both -- due to steady convection cells for up to $Ra = \mathcal{O}(10^8)$. The introduction of a mean shear flow, however, brings in additional interesting effects. The effects of both shear- and buoyancy-driven flows on the directional solidification of two-component melts have been extensively studied in the past. A detailed discussion of those studies can be found in \citet{TW2019}.

Some of the early systematic studies on shear flows over phase boundaries are those of \citet{hirata1979a, hirata1979} and \citet{Gilpin1980}. Here, we will focus on the work of \citet{Gilpin1980} because of certain features observed in their experiments. \citet{Gilpin1980} considered a turbulent boundary-layer flow over a layer of ice. At the initial instant, a groove was melted into the ice layer to introduce a perturbation at the ice-water interface. Subsequently, the effects of the shear flow on the growth of this perturbation was studied. They observed that under certain conditions, the perturbation grew and propagated downstream, leading to the formation of a ``rippled" surface. This led to an increase in the heat transfer rate by as much as 30\% - 60\% compared to a flat surface. 

\citet{Gilpin1980} attributed these observations to the effects of shear; however, because of the 4 $^{\circ}$C density maximum of water, the layer of water overlying the ice surface was unstably stratified. Hence, their observations were due to the combined effects of mean shear and buoyancy. This was recognized by \citet{TW2019}, who reanalyzed the velocity profiles from the experiments of \citet{Gilpin1980} and showed that these are described better by the Monin-Obukhov theory than the classical law of the wall \citep{Monin1971}. They also showed that the Obukhov length scale that emerged from these measurements was negative, implying the column of liquid was unstably stratified. Furthermore, \citet{TW2019} studied the stability of a phase boundary with a Rayleigh-B\'enard-Couette flow over it and showed that buoyancy destabilizes the phase boundary, whereas shear stabilizes it. They also found that for certain values of $Pe$, travelling waves are generated at the phase boundary. This tendency of buoyancy to cause large `deformations' to a phase boundary is also present in the turbulent regime: \citet{couston2020} -- who recently studied stably, neutrally, and unstably stratified shear flows over a phase boundary using direct numerical simulations (DNS) -- found that when the flow is unstably stratified, the ``channels" and ``keels" that are formed at the interface interact strongly with the underlying flow.

Here, motivated by the experiments of \citet{Gilpin1980}, we study the dynamics of an unstably stratified shear flow over a phase boundary in the laminar regime in two dimensions. Specifically, we use a combination of the Lattice Boltzmann and enthalpy methods to simulate Rayleigh-B\'enard-Poiseuille flow over a phase boundary and study their interactions. The present study is also a qualitative continuation of the work described in \citet{TW2019} into the nonlinear regime.

\section{Governing Equations}
\begin{figure}
\begin{centering}
\includegraphics[trim = 50 150 50 150, width = \linewidth]{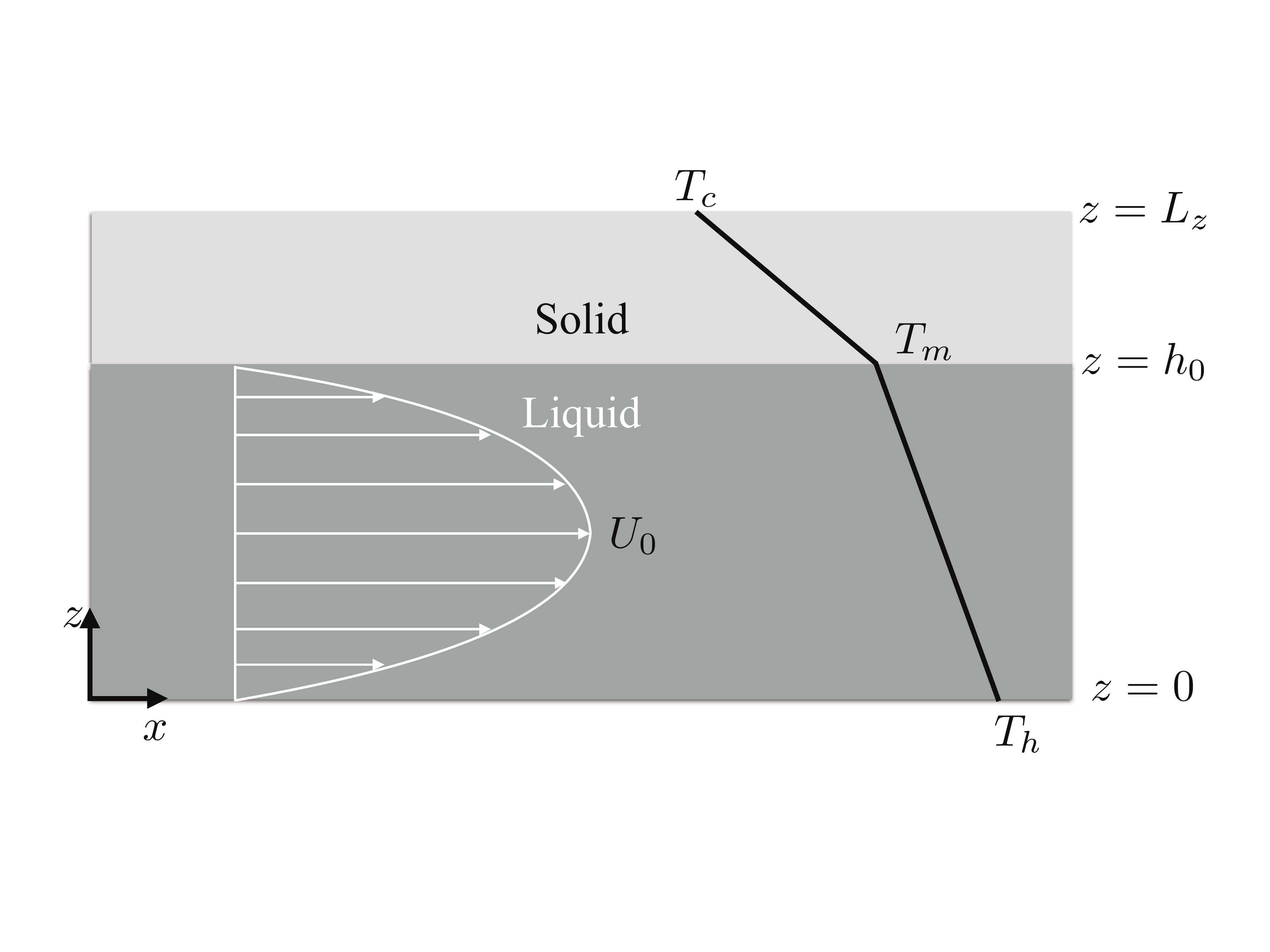} 
\caption{Schematic of the horizontally periodic domain considered here. The initial thicknesses of the liquid and solid layers are $h_0$ and $d_0 = L_z - h_0$, respectively. The temperature boundary conditions are such that $T_c < T_m < T_h$. No-slip and no-penetration boundary conditions for the velocity field are imposed at the bottom boundary and the phase boundary. The temperature fields in the liquid and solid regions at the initial instant vary only with height, and the horizontal velocity profile in the liquid region is parabolic.}
\label{fig:domain}
\end{centering}
\end{figure}

The horizontally periodic domain used in this study is shown in figure  \ref{fig:domain}. The cell height and length are $L_z$ and $L_x$, respectively. The aspect ratio of the domain is defined as $\Gamma = L_x/L_z$. Initially, the phase boundary is planar at $z = h_0$, and the fluid occupies the region $0 \le z \le h_0$. The initial thickness of the solid layer is $d_0 = L_z - h_0$.  The bottom plate is maintained at a temperature $T_h$ and the top plate is maintained at $T_c$. The melting point of the solid phase is $T_m$, and the temperature boundary conditions are such that $T_c < T_m < T_h$. We also have a fully developed Poiseuille flow in the liquid region starting from the initial instant.
As the flow develops, the initially flat phase boundary may grow/melt resulting in a deformed interface. The location of the phase boundary and the thickness of the solid layer at any time instant $t > 0$ are denoted by $h(x,t)$ and $d(x,t)$, respectively. Note that $h(x,t) + d(x,t) = L_z$.

The governing equations in the different regions are as follows.
\subsection{Liquid}
The mass, momentum, and heat balance equations are
\be
\nabla \cdot \boldsymbol{u} = 0,
\label{eqn:mass}
\ee
\be
\frac{\partial \boldsymbol{u}}{\partial t} +  \boldsymbol{u} \cdot \nabla  \boldsymbol{u} = -\frac{1}{\rho_0} \, \nabla p + g \, \alpha \, \left(T_l-T_m\right) \, \boldsymbol{\hat{z}} + \nu \, \nabla^2  \boldsymbol{u},
\label{eqn:NS_liquid}
\ee
\be
\frac{\partial T_l}{\partial t} +  \boldsymbol{u} \cdot \nabla  T_l = \kappa \, \nabla^2  T_l,
\label{eqn:heat_liquid}
\ee
respectively. Here, $\boldsymbol{u}(\boldsymbol{x},t) = (u, w)$ is the two-dimensional velocity field; $\rho_0$ is the reference density; $p(\boldsymbol{x},t)$ is the pressure field; $g$ is acceleration due to gravity; $\alpha$ is the thermal expansion coefficient; $T_l(\boldsymbol{x},t)$ is the temperature field in the liquid; $\boldsymbol{\hat{z}}$ is the unit vectors along the vertical; $\nu$ is the kinematic viscosity; and $\kappa$ is the thermal diffusivity. We assume the liquid and solid phases have the same density ($\rho_0$) and thermal diffusivity ($\kappa$).

\subsection{Solid}
The temperature field in the solid, $T_s(\boldsymbol{x},t)$, evolves according to the diffusion equation:
\be
\frac{\partial T_s}{\partial t} = \kappa \, \nabla^2  T_s.
\label{eqn:heat_solid}
\ee

\subsection{Evolution of the phase boundary}
To track the location of the phase boundary, we need an additional equation for its evolution, which is given by the Stefan condition \citep{worster2000}:
\be
\rho_0 \, L_s \, v_n =  \boldsymbol{n} \cdot \left[\boldsymbol{q_s} - \boldsymbol{q_l}\right]_{z = h}.
\label{eqn:stefan}
\ee
Here, $L_s$ is the latent heat of fusion, $v_n$ is the normal component of growth rate of the solid phase, $\boldsymbol{n}$ is the unit normal pointing into the liquid, $\boldsymbol{q_s}$ and $\boldsymbol{q_l}$ are the heat fluxes away from the interface into the solid and towards the phase boundary from the liquid, respectively.

\subsection{Boundary conditions}
We impose Dirichlet conditions on temperature at the bottom and top boundaries of the domain:
\be
T_l(z=0,t) = T_h \quad \mbox{and} \quad T_s(z = L_z,t) = T_c.
\ee
And, at the phase boundary, the temperature is the equilibrium temperature:
\be
T_l(z=h,t) = T_s(z = h,t) = T_m.
\ee

For the velocity field in the liquid region, we impose no-slip and no-penetration conditions at the bottom boundary and the phase boundary:
\be
u(z = 0, t) = w(z = 0, t) = 0;
\ee
\be
\boldsymbol{u} \cdot \boldsymbol{n} = \boldsymbol{u} \cdot \boldsymbol{t} = 0 \quad \mbox{at} \quad z = h(x,t),
\ee
where $\boldsymbol{t}$ is the unit tangent at the phase boundary. We also impose periodic boundary conditions for the temperature and velocity fields at $x = 0$ and $x = L_x$.

\subsection{Non-dimensional equations}
To non-dimensionalize the equations of motion \footnote{Except for velocity, we follow \citet{TW2019} in choosing the different scales for non-dimensionalization.}, we choose the initial centerline velocity of the Poiseuille profile in the liquid region, $U_0$, as the velocity scale; $h_0$ as the length scale, $t_0 = h_0^2/\kappa$ as the time scale, $p_0 = \rho_0 \, U_0 \, \kappa/h_0$ as the pressure scale, and $\Delta T = T_h - T_m$ as the temperature scale. Using these we obtain the dimensionless versions of equations \ref{eqn:mass}, \ref{eqn:NS_liquid}, \ref{eqn:heat_liquid}, \ref{eqn:heat_solid} and \ref{eqn:stefan} as:
\be
\nabla \cdot \boldsymbol{u} = 0;
\label{eqn:mass_scaled}
\ee
\be
\frac{\partial \boldsymbol{u}}{\partial t} +  Pe \, \left(\boldsymbol{u} \cdot \nabla  \boldsymbol{u}\right) = -\nabla p + \frac{Ra \, Pr}{Pe} \, \theta_l \, \boldsymbol{\hat{z}} + Pr \, \nabla^2  \boldsymbol{u};
\label{eqn:NS_liquid_scaled}
\ee
\be
\frac{\partial \theta_l}{\partial t} +  Pe \, \left(\boldsymbol{u} \cdot \nabla  \theta_l \right) = \nabla^2  \theta_l;
\label{eqn:heat_liquid_scaled}
\ee
\be
\frac{\partial \theta_s}{\partial t} = \nabla^2  \theta_s;
\label{eqn:heat_solid_scaled}
\ee
and
\be
v_n =  \frac{1}{\Lambda \, \mathcal{S}} \, \left[\boldsymbol{n} \cdot \left(\boldsymbol{q_s} - \boldsymbol{q_l}\right)\right]_{z = h}, 
\label{eqn:stefan_scaled}
\ee
where,
\be
\theta_l = \frac{T_l - T_m}{\Delta T} \quad \mbox{and} \quad \theta_s = \frac{T_s - T_m}{\Delta T}.
\ee
Here, we have maintained the pre-scaled notation for $\boldsymbol{u}, t$ and $\boldsymbol{x}$ for simplicity. There are five governing parameters, which are 
\be
Ra = \frac{g \, \alpha \, \Delta T \, h_0^3}{\nu \, \kappa}, \quad Pe = \frac{U_0 \, h_0}{\kappa}, \quad Pr = \frac{\nu}{\kappa},
\ee
\be
\mathcal{S} = \frac{L_s}{C_p \, \left(T_m-T_c\right)} \quad \mbox{and} \quad \Lambda = \frac{\left(T_m - T_c\right)}{\Delta T},
\ee
where $C_p$ is the specific heat of the solid phase and $\Lambda$ denotes the ratio of temperature differences in the solid and liquid regions.

The non-dimensional versions of the boundary conditions are:
\be
\theta_l(z = 0, t) = \theta_h = 1 \quad \mbox{and} \quad  \quad \theta_s(z = L_z, t) = \theta_c = -\Lambda; 
\ee
\be
\theta_s(z = h, t) = \theta_l(z = h, t) = \theta_m = 0;
\ee
\be
u(z = 0, t) =  w(z = 0, t) = 0; \quad \mbox{and} \quad 
\ee
\be
\boldsymbol{u} \cdot \boldsymbol{n} = \boldsymbol{u} \cdot \boldsymbol{t} = 0 \quad \mbox{at} \quad z = h(x,t).
\ee

\subsection{Initial conditions}
At the initial instant, the temperature profiles in the liquid and solid regions are given by:
\be
\theta_l^{(0)}(z) = 1-z,
\ee
and
\be
\theta_s^{(0)}(z) = \frac{\Lambda}{d_0} (1-z).
\ee
In addition, we demand that the heat fluxes at the phase boundary balance at the initial instant (see equation \ref{eqn:stefan_scaled}), giving
\be
\frac{d \theta_l^{(0)}}{dz} = \frac{d \theta_s^{(0)}}{dz} \quad \mbox{at} \quad  z = 1.
\ee
This gives 
\be
\Lambda = d_0.
\ee

\subsection{Heat transport}
The response of the system is quantified using the dimensionless heat flux, which is the Nusselt number, defined as
\be
Nu(t) = \left . -\frac{1}{L_x} \int_0^{L_x} \left(\frac{\partial T_l}{\partial z}\right) \, dx \middle/ \left[\frac{\Delta T}{\overline{h}(t)}\right] \right .\quad \mbox{at} \quad  z = 0.
\ee
Here, $\overline{h}(t)$ denotes the instantaneous horizontally averaged thickness of the liquid layer. After the dynamics have reached a stationary state, the horizontally and temporally averaged Nusselt number is calculated as
\be
\mathcal{N} = \frac{1}{T} \int_{t_0}^{t_0+T} \, Nu(t) \, dt.
\ee
We also define the horizontally and temporally averaged liquid height as 
\be
h_m = \frac{1}{T} \int_{t_0}^{t_0+T} \, \overline{h}(t) \, dt,
\ee
and the effective $Ra$ based on $h_m$ as
\be
Ra_e = \frac{g \, \alpha \, \Delta T \, h_m^3}{\nu \, \kappa}.
\label{eqn:Rae}
\ee
The results from this study are discussed in terms of either $Ra$ or $Ra_e$.

\section{Numerical Method}
To numerically solve the equations of motion and the boundary conditions, we combine the Lattice Boltzmann method (LBM) \citep{benzi1992, chen1998} with the enthalpy method \citep{voller1987}. In the enthalpy method, the total enthalpy is split into specific and latent heat contributions, and the regions that undergo phase change are tracked through the changes in the latent heat content of those regions \citep{voller1987}. A phase variable $\phi(\boldsymbol{x},t)$, which represents the liquid fraction field, is introduced to follow the evolution of the different phases. A grid point $\boldsymbol{x} = (x_i,z_j)$ is deemed to be solid or liquid depending on whether $\phi(\boldsymbol{x}) \le \phi_0$ or $\phi(\boldsymbol{x}) > \phi_0$, where $\phi_0 \in (0, 1)$ denotes a chosen threshold value. The choice of $\phi_0$ is arbitrary, but choosing a large value effectively increases the latent heat of fusion. This is for the following reason. The change in the nature of a grid point (solid to liquid, or \emph{vice versa}) involves a change in the latent heat of fusion. A smaller value of $\phi_0$ requires a smaller amount of heat of fusion to be provided to effect a change from solid to liquid grid point when compared with a higher value of $\phi_0$. In this study, we choose $\phi_0 = 0.5$. 

The principal advantage of the enthalpy method is that the phase boundary is not explicitly tracked, resulting in less onerous requirements for grid resolution when compared with other methods. The details of the enthalpy method can be found in \citet{voller1981} and \citet{voller1987}, and its implementation for conduction- and convection-driven phase-change problems using LBM can be found in \citet{jiaung2001} and \citet{huber2008}, respectively. For our study, we use the scheme of \citet{huber2008}. Further details are provided in Appendix \ref{sec:enthalpy}.
 
For the fluid flow, we use the D2Q9 \citep{succi2001} and D2Q5 \citep{latt} lattice models for the velocity and temperature distribution functions, respectively. No-slip and no-penetration boundary conditions for the velocity field are imposed using the mid-grid bounceback scheme \citep{succi2001}, which is known to conserve mass for flows over complex geometries in the high $Ra$ and $Re$ regimes \citep{Toppaladoddi_PhD}. The Dirichlet boundary conditions for the temperature field are imposed by requiring that the temperature distribution functions at the boundaries are the corresponding equilibrium distribution functions. 

The flow simulated by the LBM is weakly compressible, and the equation of state is the ideal gas law. Hence, it is difficult to maintain significant pressure gradients in the flow \citep{succi2001}. For these reasons, a body force $G$, which \emph{mimics} an applied pressure gradient, is introduced in the evolution equation for the velocity distribution functions. In dimensional units, the centerline velocity in plane Poiseuille flow is given by:
\be
U_0 = \frac{|\nabla p_0| \, h_0^2}{8 \, \rho_0 \, \nu},
\label{eqn:bodyforce}
\ee
where $|\nabla p_0|$ is the constant pressure gradient. Choosing a value of $U_0$, $G (= |\nabla p_0|)$ is determined using equation \ref{eqn:bodyforce}, and then used to drive the flow in the LBM. Further details on the implementation can be found in \citet{Toppaladoddi_PhD}.  

Our numerical code has been rigorously validated against spectral methods for both Rayleigh-B\'enard convection \citep{TSW2015_EPL} and Poiseuille flow \citep{TSW2015_iutam}. We have also validated the code for transient, conduction-driven melting problems against analytical solutions \citep{Toppaladoddi_PhD}. Further validation is presented in the following sections when we compare some of our results for pure convection over a phase boundary with those that exist in the literature.

\section{Results}
\subsection{Rayleigh-B\'enard convection over a phase boundary}
Here, we present results from our simulations for purely convective flow over a phase boundary. The discussion of these results serves the following two main purposes. First, it allows us to compare our results with the previous experiments and DNS studies and hence assess the accuracy of our formulation and simulation methods. And second, it provides a natural comparison point for our later discussion of the effects of mean shear on the convective motions and on the evolution of the phase boundary.

The resolution used in the simulations varies with $Ra$; e.g., for $Ra = 2.15 \times 10^3$ we use $400 \times 100$ grid points and for $Ra = 10^6$ we use $1200 \times 300$ grid points. These resolutions are such that there are at least 9 grid points in each boundary layer. Furthermore, we fix $Pr = 1$ and $h_0 = d_0 = 1$ for all simulations.

\subsubsection{Onset of thermal convection}
To study the onset of convection, we perform simulations for $Ra \in \left[1470, 1600\right]$, $\mathcal{S} = 5.82$, and $\Gamma = 10$. The value of $\mathcal{S}$ is chosen to match the experimental conditions of \citet{dietsche1985}, who used cyclohexane as the working fluid; and the large value of $\Gamma$ is chosen to ensure any finite-size effects are minimized. The $Pr$ for cyclohexane is $17.6$ \citep{dietsche1985}, but we use $Pr = 1$ in our simulations. This choice does not affect the onset of convection as $Ra_c$ is independent of $Pr$ for this system \citep{davis1984, TW2019}.

\begin{figure}
\begin{centering}
\includegraphics[trim = 0 0 0 0, width = \linewidth]{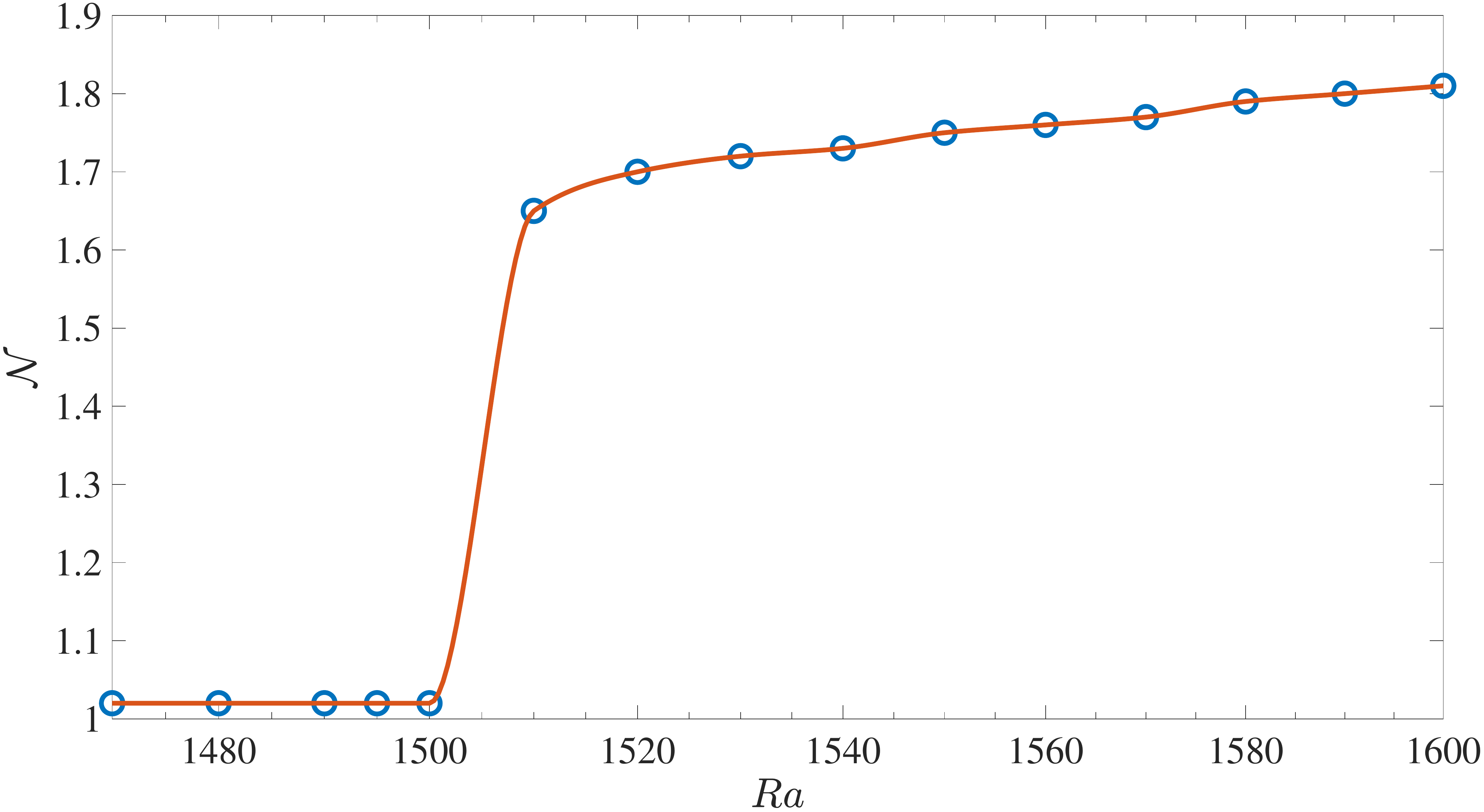} 
\caption{$\mathcal{N}(Ra)$ for $Ra \in \left[1470, 1600\right]$, $Pr =1$, and $\mathcal{S} = 5.82$. The critical $Ra$ is $Ra_c \approx 1510$.}
\label{fig:NuRa_davis}
\end{centering}
\end{figure}

Figure \ref{fig:NuRa_davis} shows $\mathcal{N}(Ra)$ for $Ra \in \left[1470, 1600\right]$. There is a jump in $\mathcal{N}$ at the onset of convection, which is at $Ra = 1510$. This behaviour is in contrast to what happens near $Ra_c$ in the classical Rayleigh-B\'enard convection (RBC) \citep{chandra2013}, and is in good agreement with the theoretical prediction of \citet{davis1984}. Similar behaviour near $Ra = Ra_c$ has been reported in previous experiments \citep{dietsche1985} and DNS studies \citep{esfahani2018, purseed2020}. 

\begin{figure}
\begin{centering}
\includegraphics[trim = 0 0 0 0, width = \linewidth]{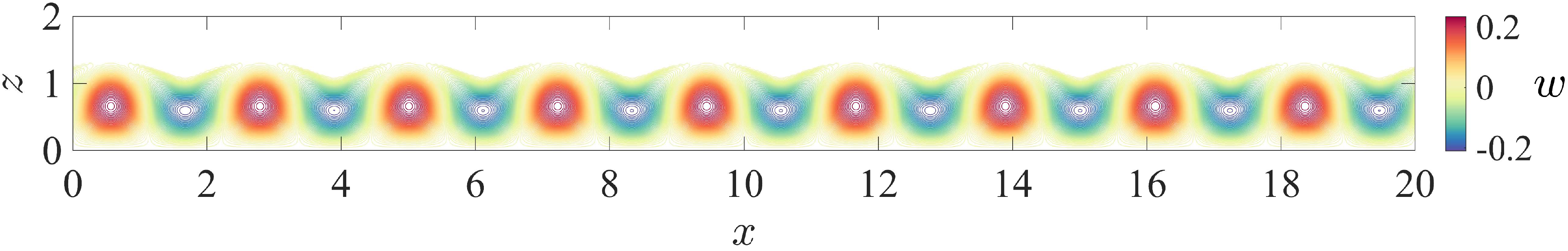} 
\caption{Contours of steady state vertical velocity field for $Ra = 1510$. The solid phase is shown in white. The velocity is non-dimensionalized by the buoyancy velocity scale $\sqrt{g \, \alpha \, \Delta T \, h_0}$.} 
\label{fig:vel3} 
\end{centering}               
\end{figure}

Figure \ref{fig:vel3} shows the contours of steady state vertical velocity field for $Ra = 1510$.
We can calculate the critical wavenumber from figure \ref{fig:vel3}, noting that there are nine pairs of counter-rotating cells. This gives the dimensionless wavelength as $\lambda = 20/9 \approx 2.22$ and the critical wavenumber as $k_c = 2 \pi/\lambda \approx 2.83$. These values are in excellent agreement with $Ra_c = 1493$ and $k_c = 2.82$ from the linear stability calculations of \citet{davis1984}.

\subsubsection{Thermal convection for larger $Ra$}
Before exploring the combined effects of shear and buoyancy on the evolution of the phase boundary, we investigate the effects of pure thermal convection for $Ra \in \left[2.15 \times 10^3, 10^6\right]$. The simulation results reported in the remainder of this paper are for $\mathcal{S} = 1$, except in the last subsection, and $\Gamma = 4$.

\begin{figure}
\begin{centering}
\includegraphics[trim = 0 0 0 0, width = \linewidth]{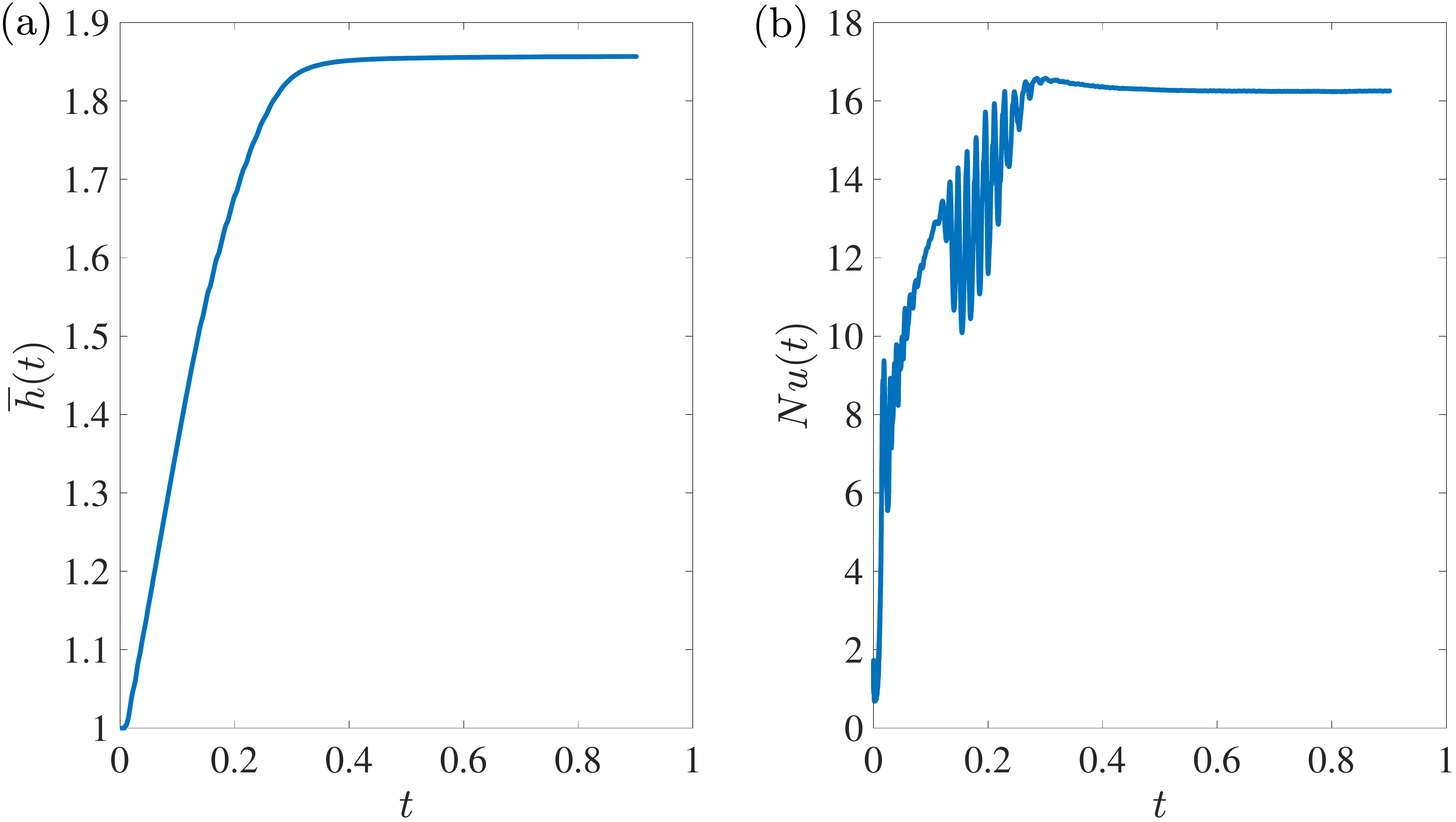} 
\caption{Time series for the horizontally averaged (a) height of the liquid column and (b) heat flux for $Ra = 10^6$.}
\label{fig:timeseries}
\end{centering}
\end{figure}

\begin{figure}
\begin{centering}
\includegraphics[trim = 0 0 0 0, width = \linewidth]{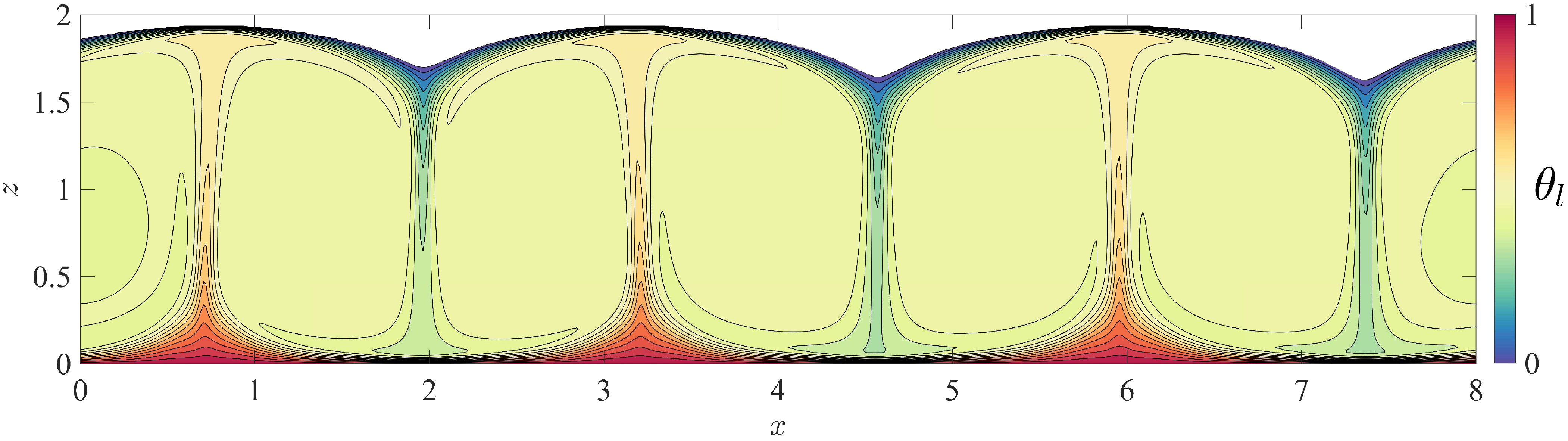} 
\caption{Snapshot of the steady temperature field for $Ra = 10^6$.}
\label{fig:temp_RBC}
\end{centering}
\end{figure}

In figures \ref{fig:timeseries}(a) and \ref{fig:timeseries}(b) we show the time series for the horizontally averaged thickness of the liquid layer and the heat flux for $Ra = 10^6$. The following observations can be made from these figures: (1) after an initial transient, both the liquid height and the heat flux attain steady state; (2) the $Nu(t)$ time series exhibits oscillations before reaching the steady state. These oscillations are due to the evolving convection cells, whose aspect ratio continuously changes before reaching the steady-state value. The effective $Ra$ for this case is $Ra_e \approx 6.5 \times 10^6$ and the steady state $\mathcal{N} = 16.27$, which is larger than $\mathcal{N} = 12.07$ for classical RBC \citep{doering2009}. These results are in qualitative agreement with the findings of \citet{favier2019} and \citet{purseed2020}.
The increase in the heat flux compared to classical RBC is because the non-planar phase boundary ``locks in" the convection cells, thereby delaying the onset of unsteady convection \citep{favier2019}. This is seen in figure \ref{fig:temp_RBC}, which shows a snapshot of the steady temperature field for $Ra = 10^6$. A close examination of the cusps at the phase boundary in figure \ref{fig:temp_RBC} reveals that they have slightly different amplitudes.

\begin{figure}
\begin{centering}
\includegraphics[trim = 0 0 0 0, width = \linewidth]{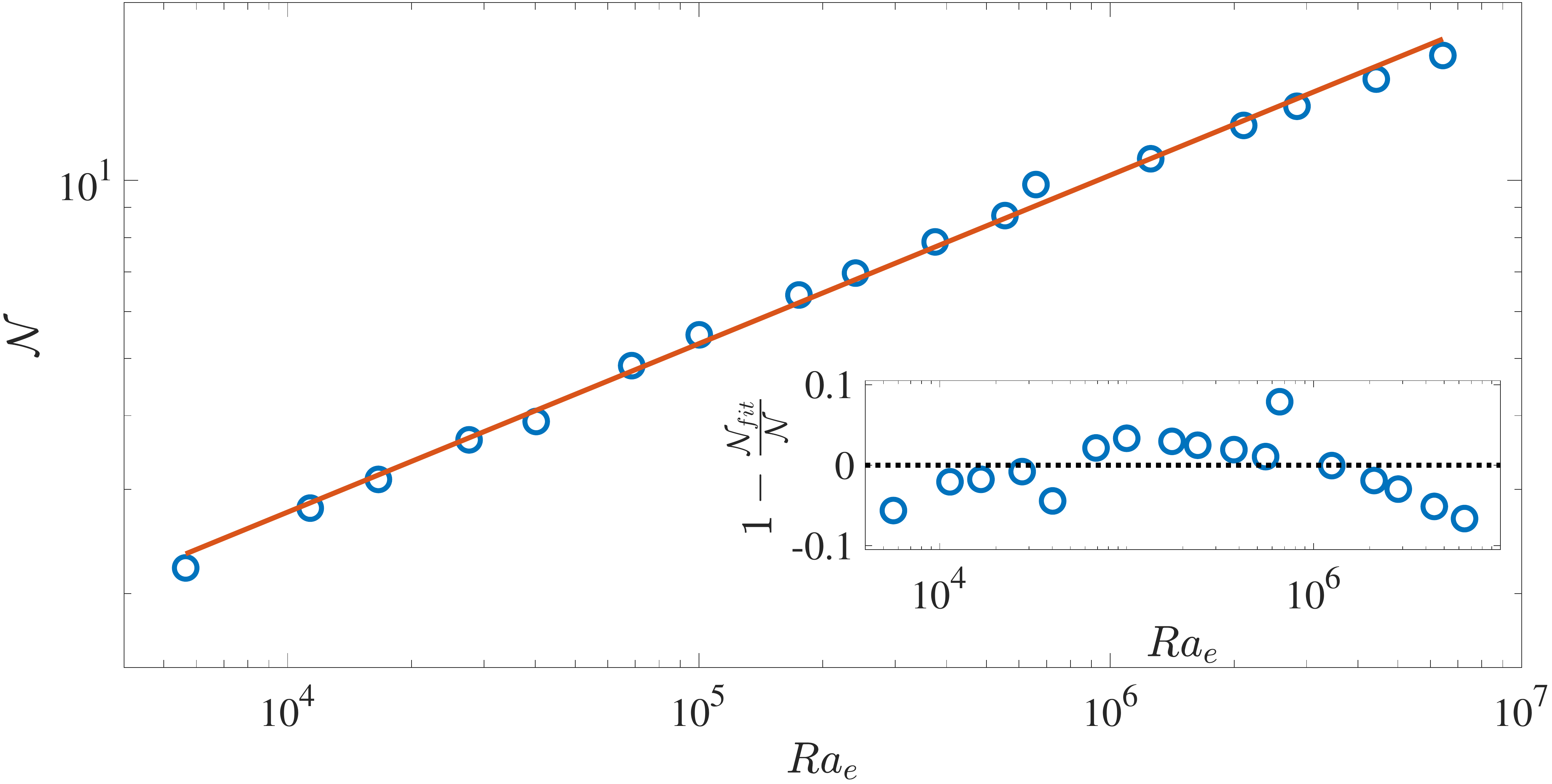} 
\caption{$\mathcal{N}$ as a function of $Ra_e$. The latter is calculated using equation \ref{eqn:Rae}. Symbols are data from simulations and the solid line is the fit $Nu = 0.2 \times Ra_e^{0.285 \pm 0.009}$. The error bars on the exponent represent the 95\% confidence interval. The inset shows the residuals from the fit. The curvature in the residual indicates that there is a weak deviation from the power-law fit.}
\label{fig:NuRa_RBC}
\end{centering}
\end{figure}

To understand the impact of the phase boundary on the dependence of heat flux on buoyancy forcing, we plot $\mathcal{N}$ as a function of $Ra_e$ in figure \ref{fig:NuRa_RBC}. The data are described well by the power law $\mathcal{N} = 0.2 \times Ra_e^{0.285 \pm 0.009}$, which is obtained from a linear least-squares fit to the $\log \mathcal{N} - \log Ra_e$ data. The exponent $\beta = 0.285$, which is indistinguishable from $\beta = 2/7$, is in remarkable agreement with the findings of previous DNS studies of classical RBC \citep{doering2009, TSW2015_EPL}. However, the prefactor here is larger than that in the classical RBC case. This is because it depends on the geometry of the boundaries \citep{TSW2015_EPL}. This effect on the prefactor has been reported by \citet{favier2019} as well, and they obtained $\beta \approx 0.27$.

Another feature that is absent in figure \ref{fig:NuRa_RBC} is a discontinuity in the $\mathcal{N}(Ra_e)$ data at around $Ra_e = 10^6$, which is due to a pattern competition instability observed in the classical RBC \citep{glazier1999, doering2009}. This indicates that the phase boundary suppresses this instability. However, this does not rule out its appearance at a higher $Ra$.

\begin{figure}
\begin{centering}
\includegraphics[trim = 0 0 0 0, width = \linewidth]{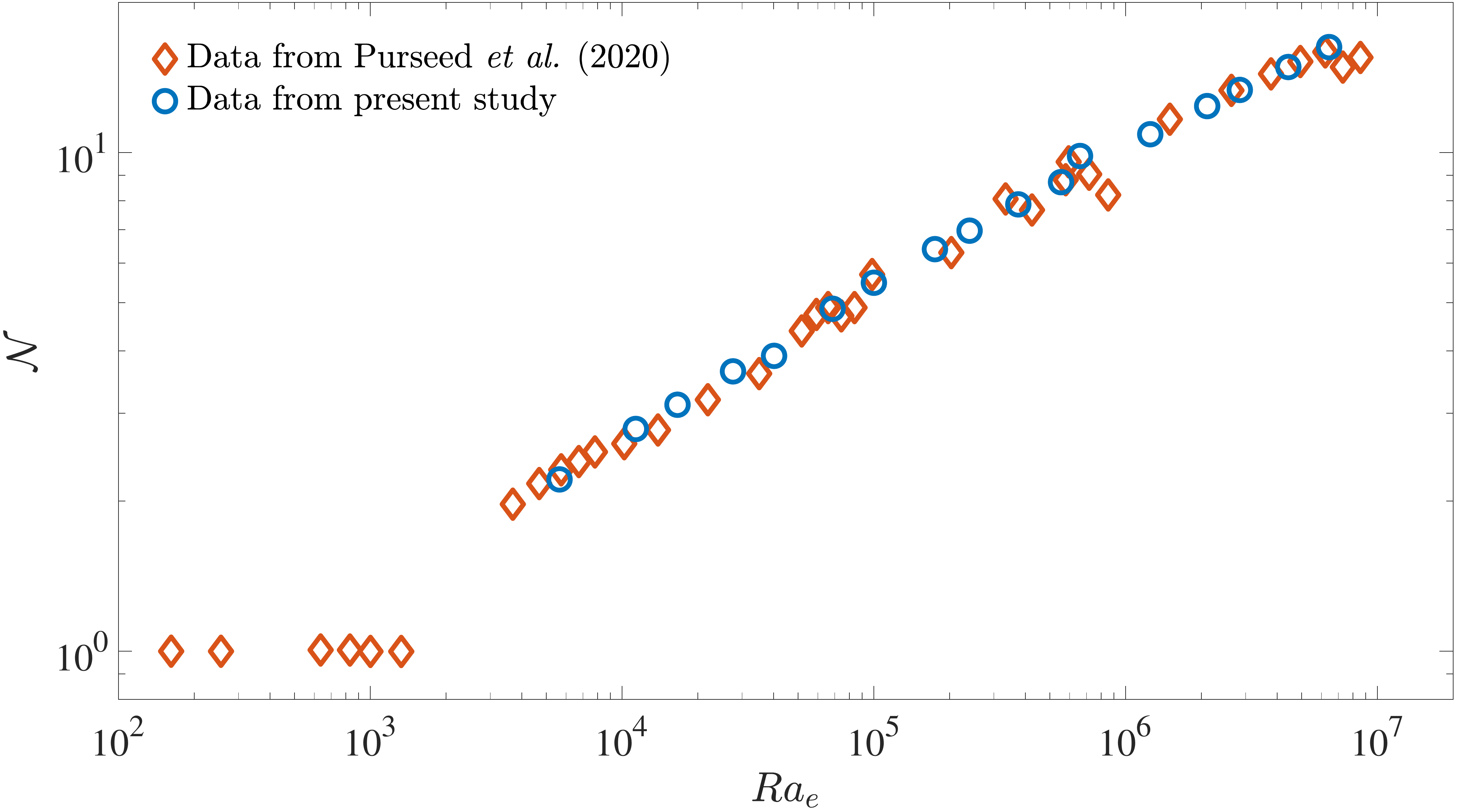} 
\caption{Comparison of $\mathcal{N}(Ra_e)$ data with those from \citet{purseed2020}, who had $\Gamma = 6$, $h_0 = 1.8$ and $\mathcal{S} = Pr = 1$ in their simulations. Circles are data points from the present study and diamonds are from \citet{purseed2020}.}
\label{fig:NuRa_favier}
\end{centering}
\end{figure}

In figure \ref{fig:NuRa_favier}, we show our $\mathcal{N}(Ra_e)$ data along with those from \citet{purseed2020}, who had $\Gamma = 6$, $h_0 = 1.8$ and $\mathcal{S} = Pr = 1$ in their simulations. The agreement between the results shows that for a fixed $Pr$, $\mathcal{N}$ depends only on $Ra_e$ and does not appreciably depend on the initial conditions. 

\subsection{Rayleigh-B\'enard-Poiseuille flow over a phase boundary}
Having established consistency of our simulations with previous work on coupled convection and phase change, we now explore the effects of mean shear on both the convective motions and the evolution of the phase boundary. The range of $Pe$ used in this study is $Pe \in [0,50]$. The simulations of Rayleigh-B\'enard-Poiseuille flow are equally well resolved as our simulations of RBC over phase boundary, with at least 9 grid points in each boundary layer.

\subsubsection{Mean height of the liquid layer}
\begin{figure}
\begin{centering}
\includegraphics[trim = 0 0 0 0, width = \linewidth]{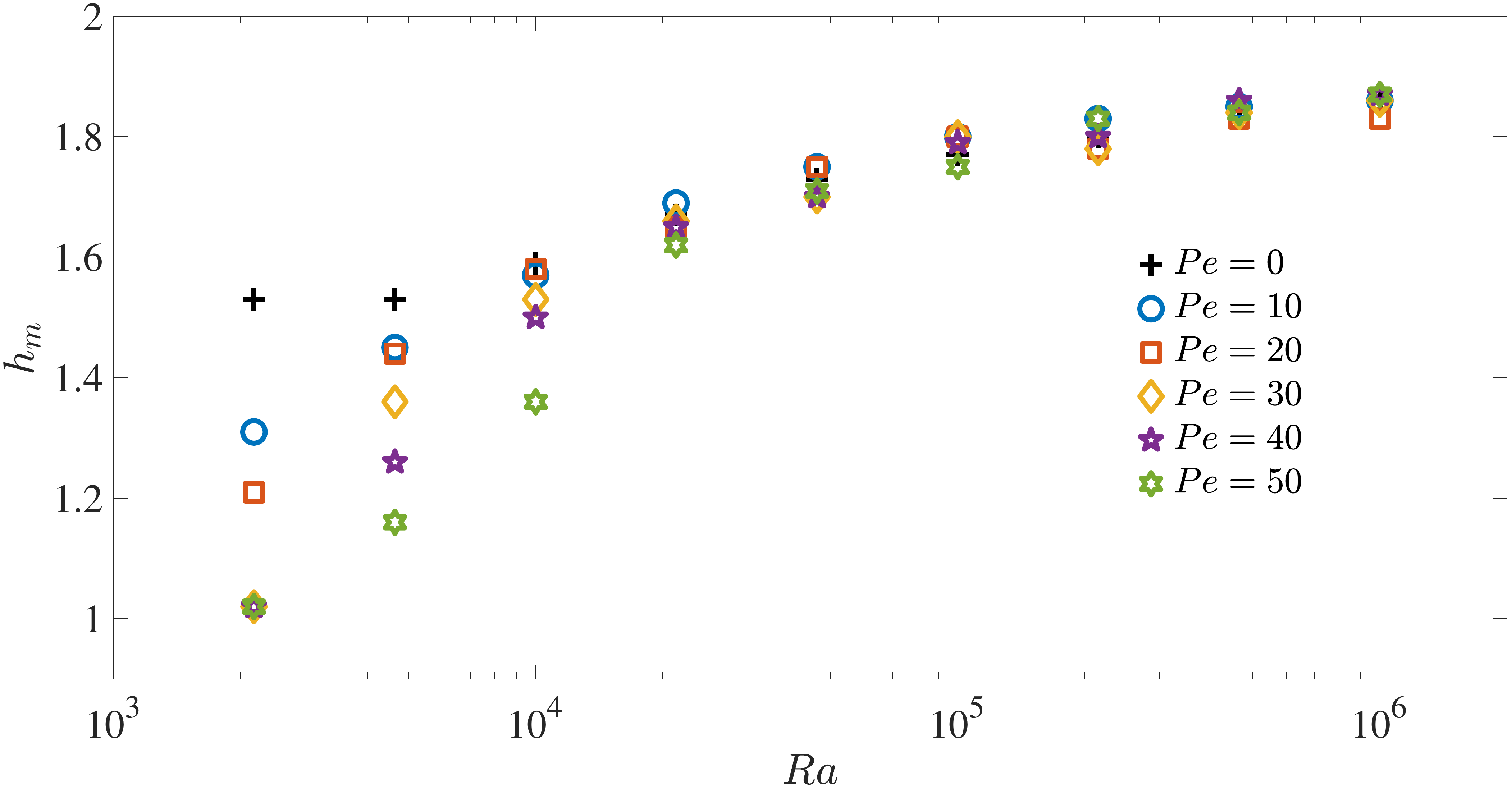} 
\caption{Mean height of the liquid layer, $h_m$, as a function of $Ra$ for the different values of $Pe$.}
\label{fig:height_Ra}
\end{centering}
\end{figure}

We first consider the combined effects of mean shear and buoyancy on the mean height of the liquid layer. In figure \ref{fig:height_Ra}, $h_m$ is shown as a function of $Ra$ for the different $Pe$ considered.
The following observations can be made from the figure: (i) with increasing $Ra$, the variation in $h_m$ for the different $Pe$ decreases; (ii) for $Pe = 40$ and $50$ and the lowest $Ra$, there is negligible melting of the phase boundary, indicating there is no bifurcation to steady convection; and (iii) for a fixed $Ra \ge 2.15 \times 10^4$, the changes in $h_m$ are not monotonic with $Pe$. These observations indicate that the interplay between the shear flow and convection has substantial effects on the evolution of the phase boundary.

\subsubsection{Heat transport}
\begin{figure}
\centering

\begin{subfigure}
\centering
\includegraphics[trim = 0 0 0 0, clip, width = \linewidth]{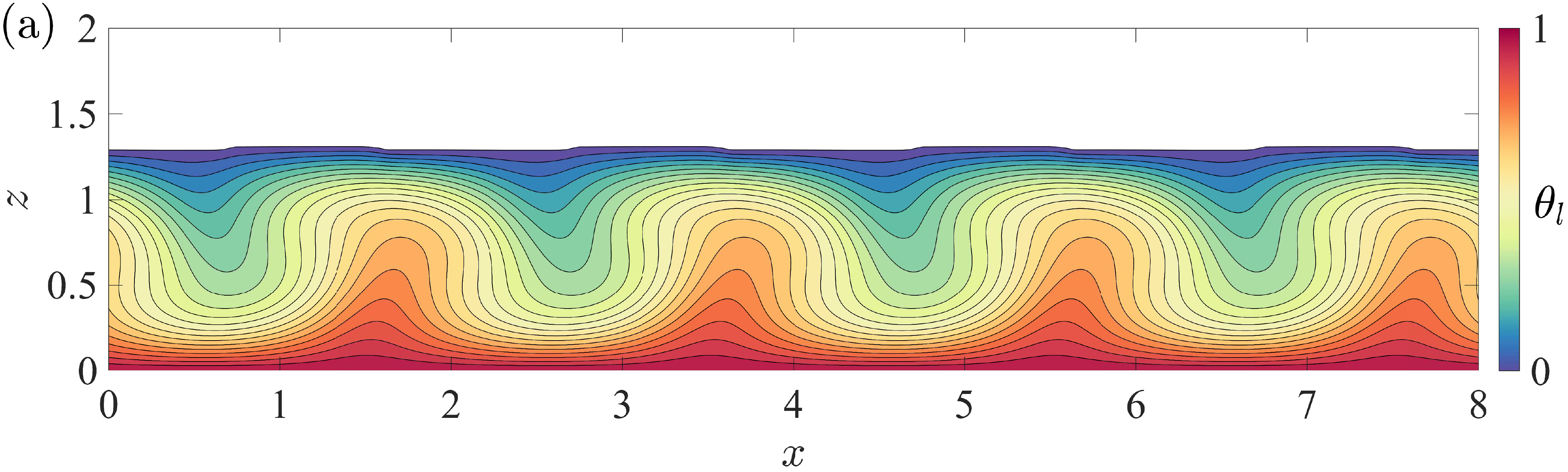}  
\end{subfigure}
    
\begin{subfigure}
\centering
\includegraphics[trim = 0 0 0 0, clip, width = \linewidth]{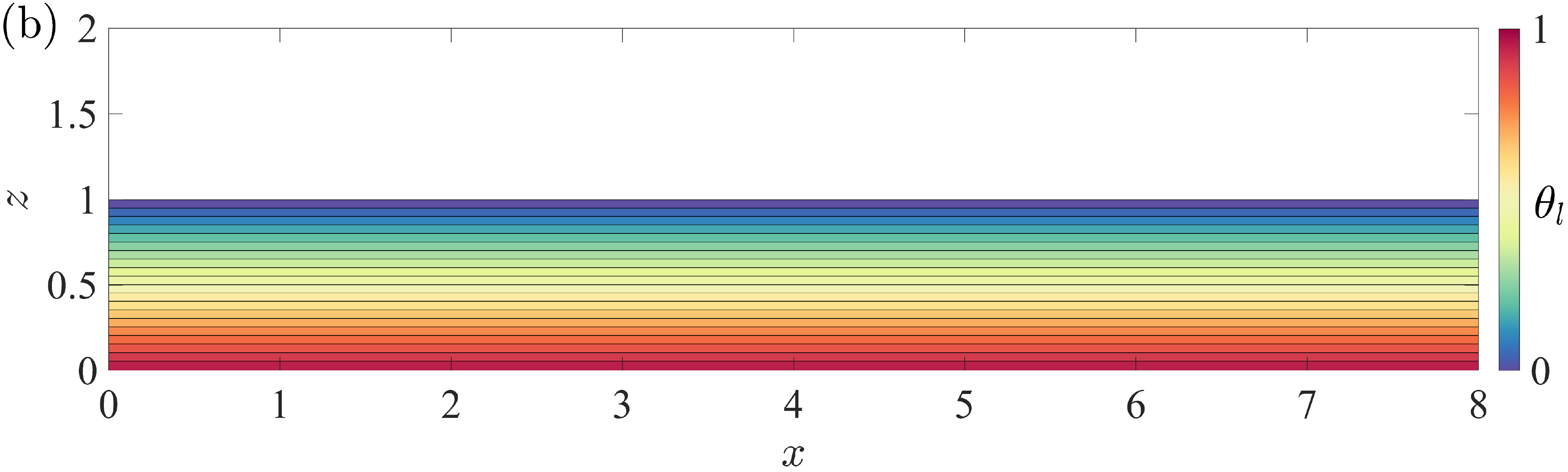} 
\end{subfigure}
                    
\caption{Temperature fields for $Ra = 2.15 \times 10^3$ and (a) $Pe = 10$ and (b) $Pe = 50$ at $t = 49.84$. Convective motions are suppressed for $Pe = 50$.}     
\label{fig:temp_Re}             
\end{figure}

To understand these effects, we consider the impact of mean shear and buoyancy on the transport of heat. In figure \ref{fig:temp_Re} we show the temperature fields for $Ra = 2.15 \times 10^3$ and (a) $Pe = 10$ and (b) $Pe = 50$ at $t = 49.84$. The deformation of the phase boundary in figure \ref{fig:temp_Re}(a) is due to the convective cells. The mean shear flow has a considerable effect on the convective motions: for $Pe = 10$ the convection cells are slightly distorted, but for $Pe = 50$ the convective motions disappear completely. 

\begin{figure}
\begin{centering}
\includegraphics[trim = 0 0 0 0, width = \linewidth]{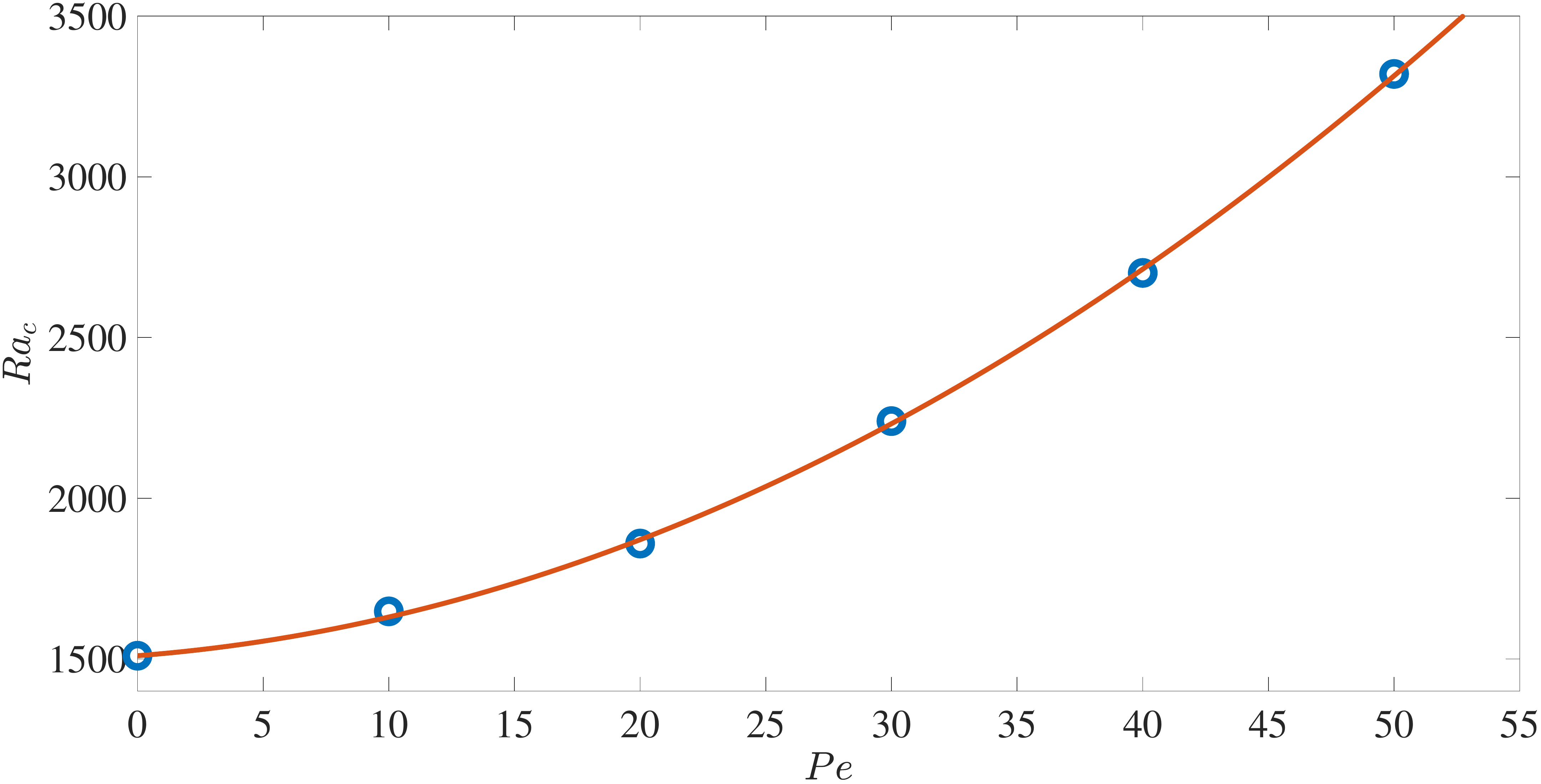} 
\caption{Critical Rayleigh number for the onset of convection as a function of $Pe$. The circles are data from simulations, and the solid line is the quadratic fit. For $Pe = 0$, $\Gamma = 10$; and, for $Pe > 0$, $\Gamma = 4$.}
\label{fig:onset}
\end{centering}
\end{figure}

\begin{figure}
\begin{centering}
\includegraphics[trim = 0 0 0 0, width = \linewidth]{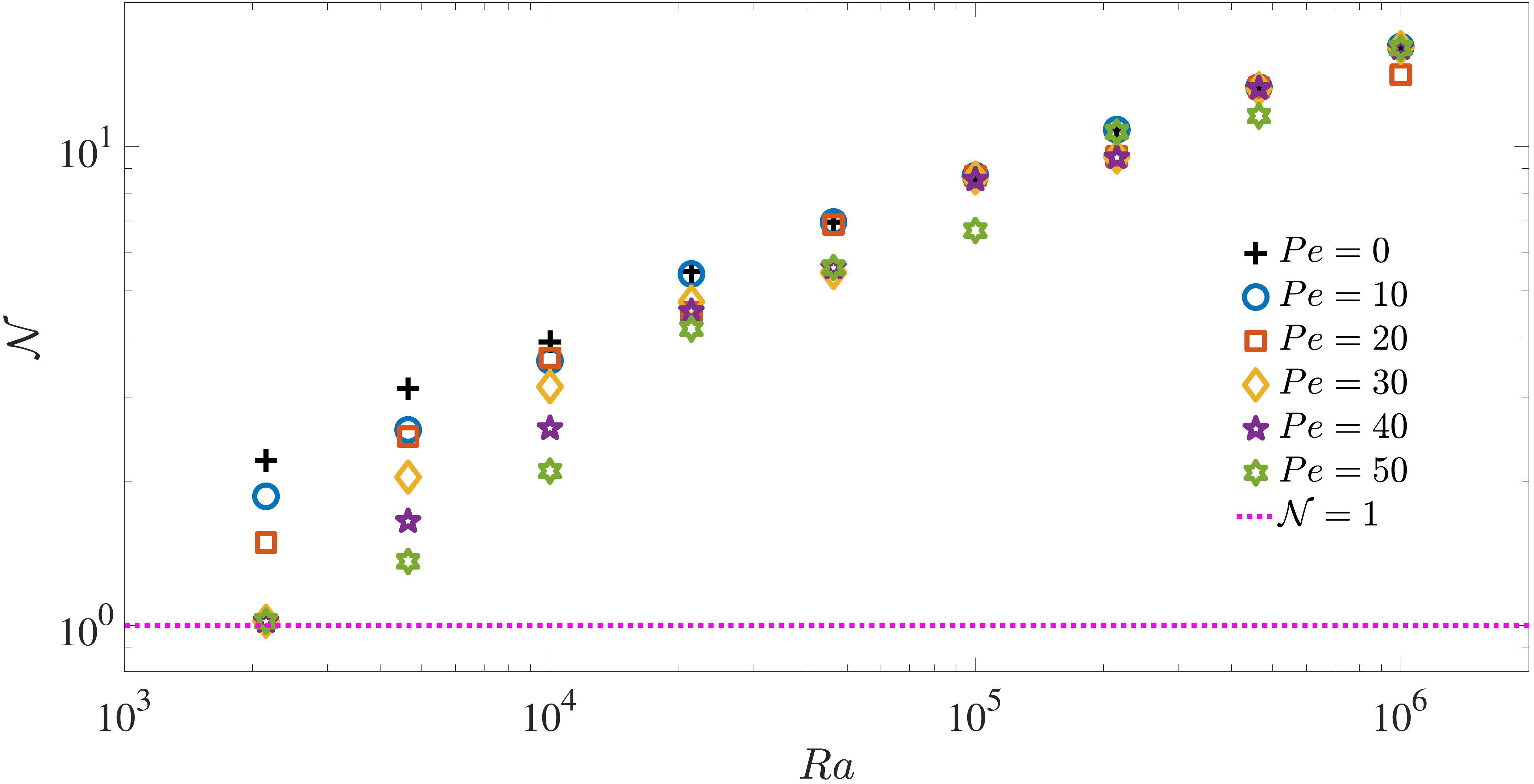} 
\caption{$\mathcal{N}$ as a function of $Ra$ for the different $Pe$.}
\label{fig:NuRaPe}
\end{centering}
\end{figure}

The effects of the mean shear on convective motion can be seen more clearly by considering its effects on $Ra_c$, which is shown in figure \ref{fig:onset}, and on the heat transport, which is shown in figure \ref{fig:NuRaPe}. The $Ra_c$ is a monotonically increasing function of $Pe$; the solid line in figure \ref{fig:onset} shows the quadratic fit to the data.
The behaviour of $\mathcal{N}$ with $Ra$ and $Pe$ in figure \ref{fig:NuRaPe} is qualitatively similar to that of $h_m$ (figure \ref{fig:height_Ra}). To obtain a more complete understanding, the relative effects of mean shear and buoyancy have to be considered.

\begin{figure}
\begin{centering}
\includegraphics[trim = 0 0 0 0, width = \linewidth]{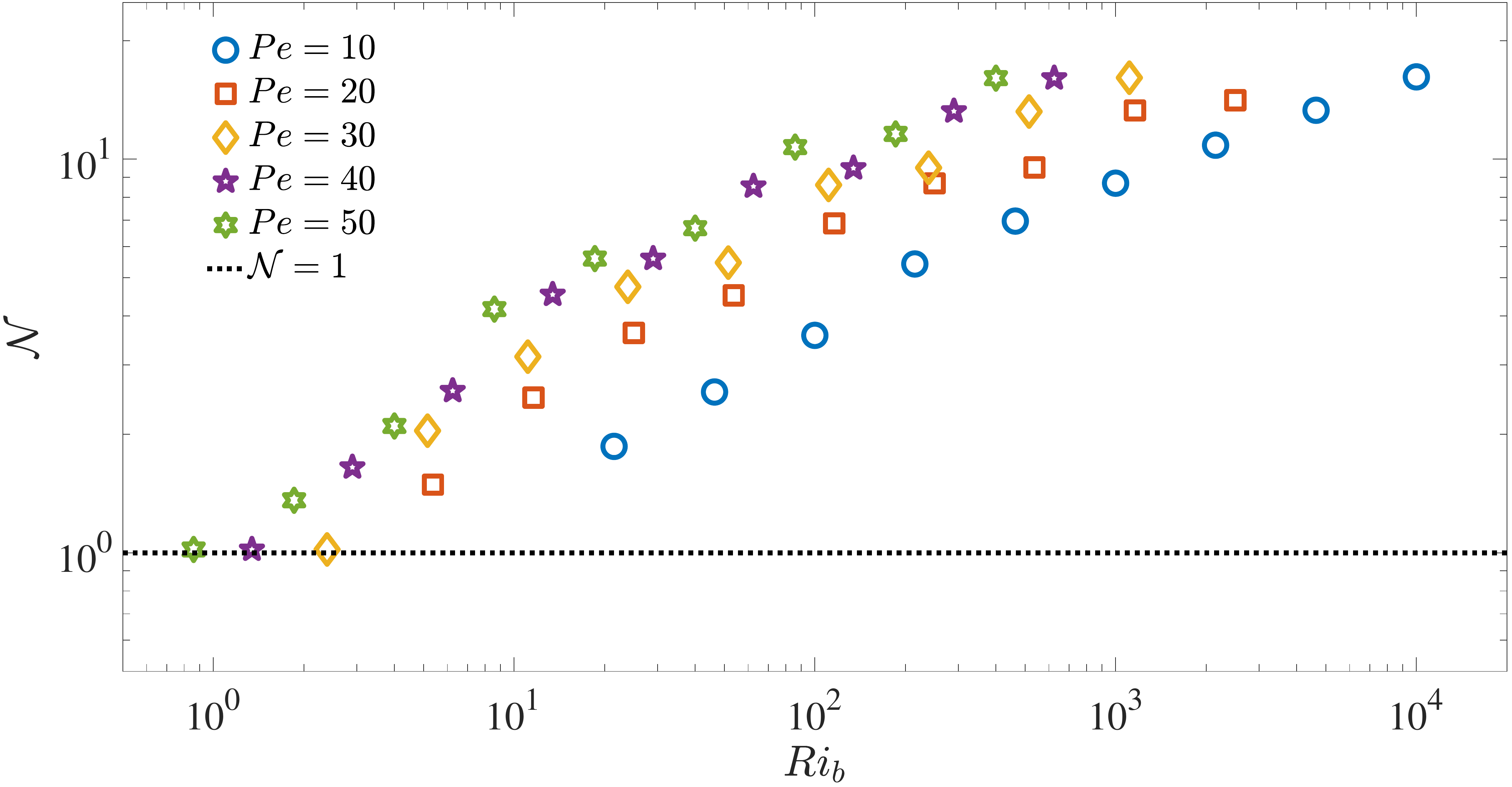} 
\caption{$\mathcal{N}$ as a function of $Ri_b$ for the different $Pe$. For each $Pe$, the simulations cover the $Ra$ range $Ra \in \left[2.15 \times 10^3, 10^6\right]$.}
\label{fig:NuRi}
\end{centering}
\end{figure}

To quantify the relative strengths of buoyancy and mean shear, we introduce a bulk Richardson number, defined as \citep{chandra2013}
\be
Ri_b = \frac{g \, \alpha \, \Delta T \, h_0}{U_0^2} = \frac{Ra \cdot Pr}{Pe^2},
\ee
and use it to study the changes in $\mathcal{N}$ for different values of $Ra$ and $Pe$. In figure \ref{fig:NuRi} we show the dependence of $\mathcal{N}$ on $Ri_b$ for the different $Pe > 0$. For $Ri_b = \mathcal{O}(1)$, the mean shear dominates and hence the heat transport is only due to conduction. However, for $Ri_b \gg 1$ buoyancy dominates and the values of $\mathcal{N}$ are close to those for purely convective flow (see figure \ref{fig:NuRaPe}). For a fixed value of $Ra$, $\mathcal{N}$ does not increase monotonically with decreasing $Pe$ because the changes in the value of $h_m$ and, hence, $Ra_e$ are not monotonic with $Pe$.

\begin{figure}
\begin{centering}
\includegraphics[trim = 0 0 0 0, width = \linewidth]{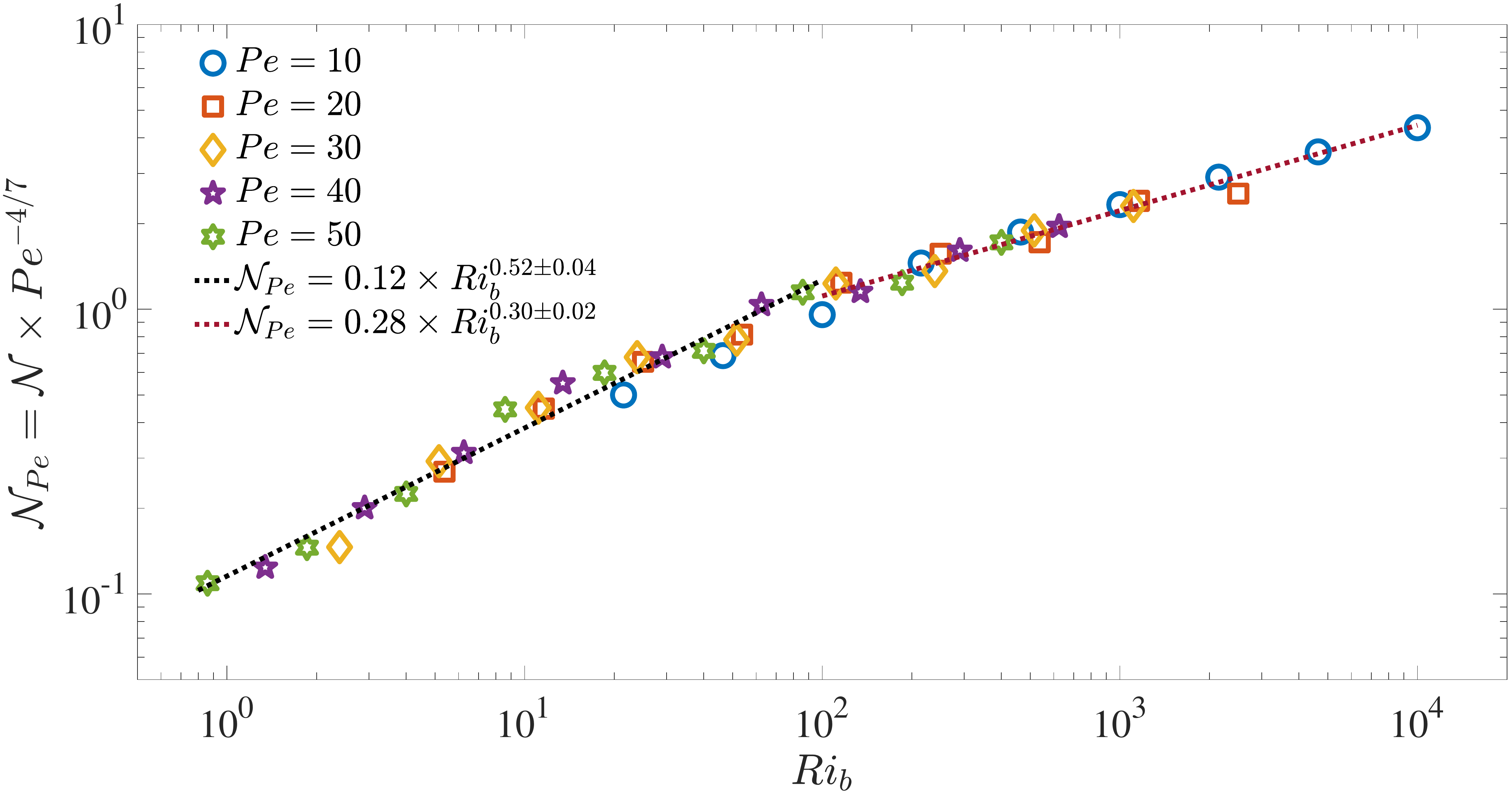} 
\caption{$\mathcal{N}_{Pe} = \mathcal{N} \times Pe^{-4/7}$ as a function of $Ri_b$. The $\mathcal{N}(Pe, Ri_b)$ data sets shown in figure \ref{fig:NuRi} collapse for this scaling.}
\label{fig:NuRiCollapse}
\end{centering}
\end{figure}

In order to determine $\mathcal{N} = \mathcal{N}(Pe, Ri_b)$, we assume that this functional relation is of the form
\be
\mathcal{N} = A \, Pe^{\gamma_1} \, Ri_b^{\gamma_2},
\label{eqn:func1}
\ee
where $A, \gamma_1, \gamma_2 > 0$. Writing equation \ref{eqn:func1} in terms of $Pe$ and $Ra$, we have
\be
\mathcal{N} = A \, Pe^{\gamma_1 - 2 \, \gamma_2} \, Ra^{\gamma_2}.
\ee
In the limit $Ri_b \rightarrow \infty$ and $Pe = \mathcal{O}(1)$, we expect the mean shear to play no role in heat transport; hence, we should recover the $\mathcal{N}-Ra$ scaling law for pure convection. This leads to $\gamma_2 = 2/7$ and $\gamma_1 - 2 \, \gamma_2 = 0$, giving $\gamma_1 = 4/7$. Hence, from equation \ref{eqn:func1} we get
\be
\frac{\mathcal{N}}{Pe^{4/7}} = \mathcal{F}(Ri_b),
\ee
where $\mathcal{F}$ is a power-law function of $Ri_b$. In figure \ref{fig:NuRiCollapse} we plot $\mathcal{N}_{Pe} = \mathcal{N} \times Pe^{-4/7}$ vs. $Ri_b$, and observe that this rescaling achieves a collapse of the different data sets shown in figure \ref{fig:NuRi}.
The collapsed data set can be described using two power laws, which are obtained from the linear least-squares fits to $\log \mathcal{N}_{Pe}-\log Ri_b$ data:
\be
\mathcal{N}_{Pe} = 0.12 \times Ri_b^{0.52 \pm 0.04}
\ee
for $Ri_b \in \left[0.86, 100\right]$ and
\be
\mathcal{N}_{Pe} = 0.28 \times Ri_b^{0.30 \pm 0.02}
\ee
for $Ri_b \in \left[100, 10000\right]$. The mean shear is found to appreciably affect the convective flow dynamics up to $Ri_b = \mathcal{O}(100)$ (see figure \ref{fig:phase}); hence, the segmentation of the $\mathcal{N}_{Pe}(Ri_b)$ data set for determining the power laws. The exponent of the second power law is close to $\gamma_2 = 2/7$, with the small difference indicating a weak influence of the mean shear on the heat transport.

\subsubsection{Pattern competition} \label{sec:pattern}
\begin{figure}
\begin{centering}
\includegraphics[trim = 0 0 0 0, width = \linewidth]{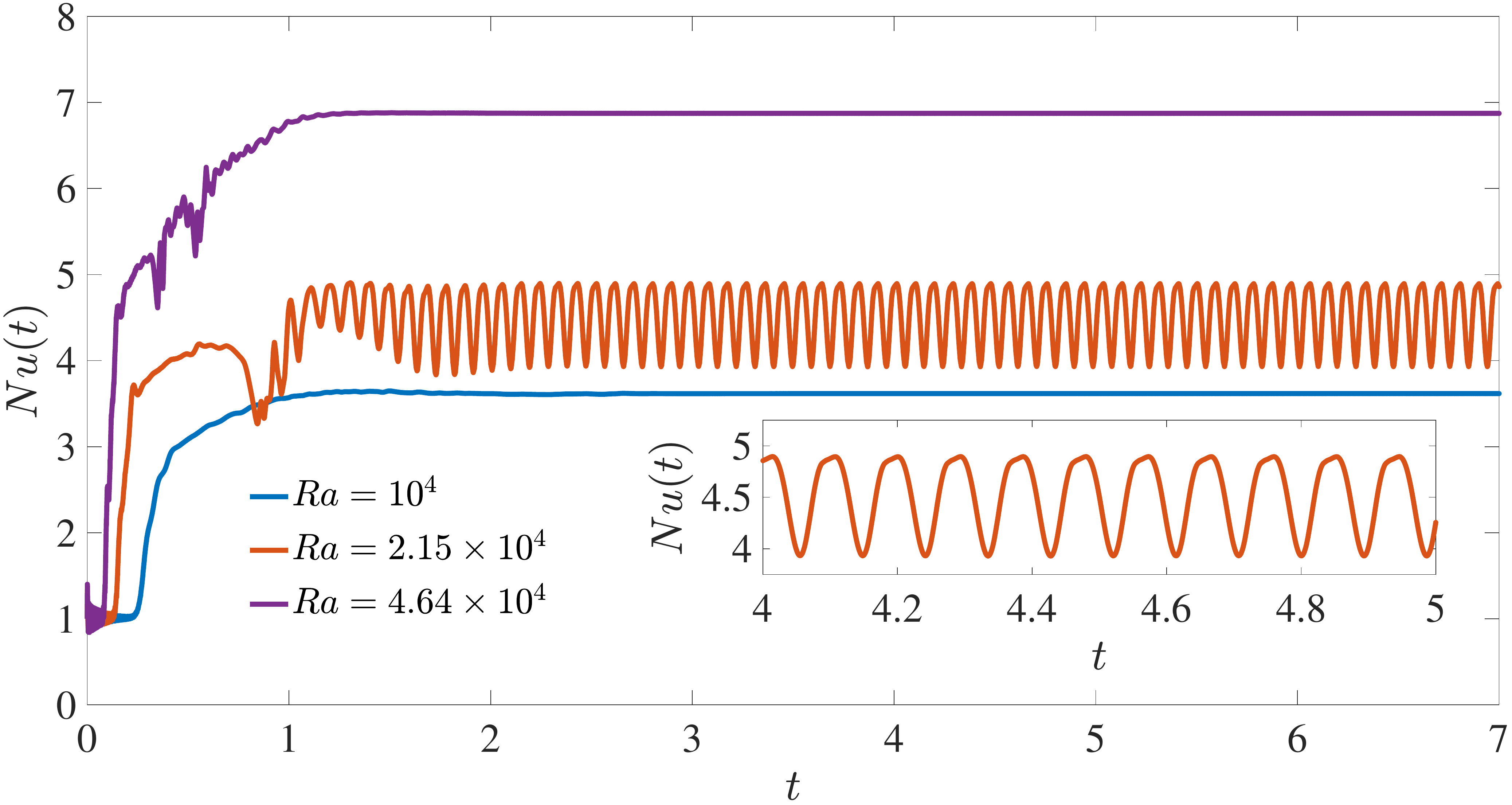} 
\caption{Time series of horizontally averaged heat flux, $Nu(t)$, for $Pe = 20$ and $Ra = 10^4, 2.15\times 10^4$, and $4.64 \times 10^4$. The inset shows the oscillations for $Ra = 2.15 \times 10^4$.}
\label{fig:pattern_time}
\end{centering}
\end{figure}

For the range of $Ra$ and $Pe$, and hence $Ri_b$, studied here, the heat flux reaches a steady value for $Ri_b = \mathcal{O}(1)$ and $Ri_b \gg 1$. However, for certain intermediate values of $Pe$ and $Ra$, it becomes periodic. These values of $Pe$ and $Ra$ correspond to $Ri_b \in \left[15, 95\right]$. In figure \ref{fig:pattern_time}, we show the $Nu(t)$ time series for $Pe = 20$ and $Ra = 10^4, 2.15\times 10^4$, and $4.64 \times 10^4$. The heat transport becomes steady for the lowest and highest $Ra$ here, but attains a periodic state for $Ra = 2.15 \times 10^4$.

\begin{figure}
\begin{centering}
\includegraphics[trim = 0 0 0 0, width = \linewidth]{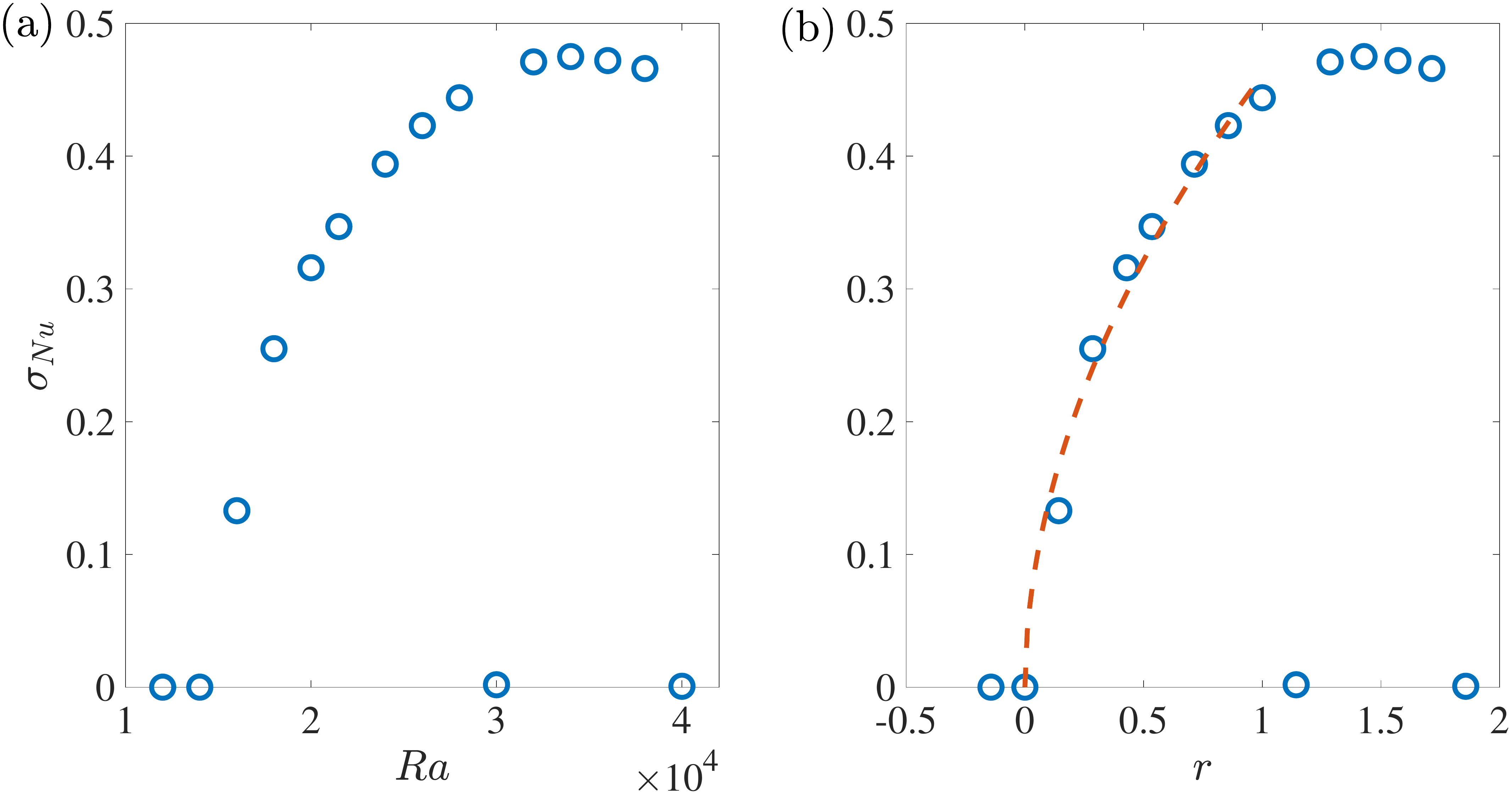} 
\caption{Bifurcation diagram for $Pe = 20$ and $Ra \in \left[1.2\times10^4, 4 \times 10^4\right]$. In figure (a), the standard deviation of the $Nu(t)$ time series, $\sigma_{Nu}$, is plotted as function of $Ra$, and in (b) $\sigma_{Nu}$ is plotted as a function of $r = (Ra-Ra_1)/Ra_1$, where $Ra_1$ denotes the Rayleigh number at the bifurcation point and is $1.4 \times 10^4$ in this case. The circles are data points from simulations and the dashed line in (b) is the fit $\sigma_{Nu} = 0.46 \times r^{0.52 \pm 0.12}$.}
\label{fig:sigma1}
\end{centering}
\end{figure}

In order to understand this behaviour in the neighbourhood of $Pe = 20$ and $Ra = 2.15 \times 10^4$, we perform additional simulations for $Ra \in \left[1.2\times10^4, 4 \times 10^4\right]$. The amplitude of the oscillations is quantified using the standard deviation of the $Nu(t)$ time series, $\sigma_{Nu}$. Figure \ref{fig:sigma1}(a) shows the bifurcation diagram in this neighbourhood. 
We see that the oscillations in $Nu(t)$ first occur at $Ra = 1.6 \times10^4$, reaching their maximum amplitude at $Ra = 3.4 \times 10^4$, and finally vanishing at $Ra = 4 \times 10^4$. The oscillations also vanish at $Ra = 3 \times 10^4$, where the heat flux reaches a steady state. These windows of periodic states are reminiscent of the window of ``self-oscillations" that is observed in the dynamics of the Sel'Kov oscillator, which is a simplified mathematical model of glycolysis, for certain range of its parameter values \citep{sel1968, strogatz2018}.

\begin{figure}
\begin{centering}
\includegraphics[trim = 0 0 0 0, width = \linewidth]{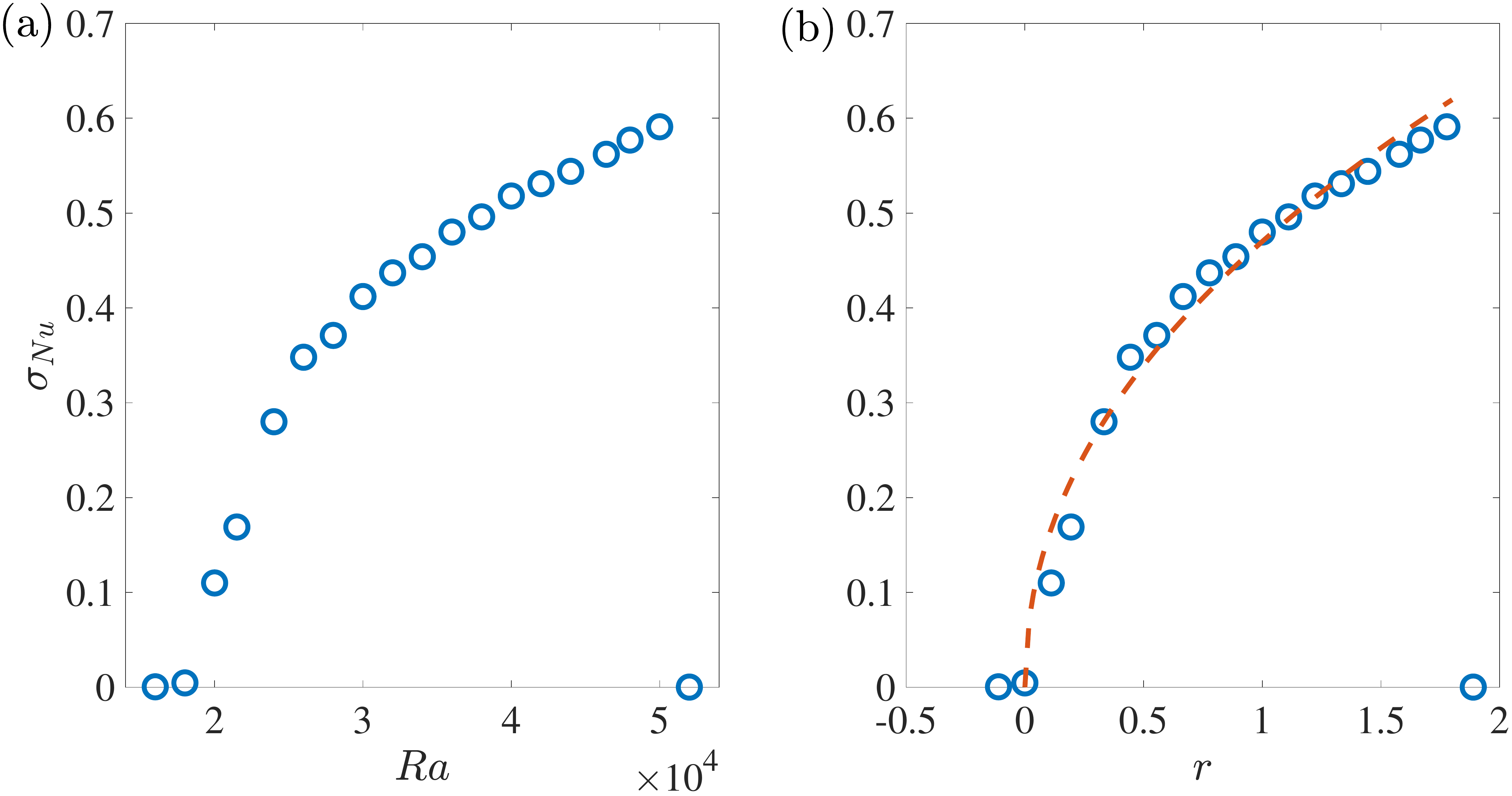} 
\caption{Bifurcation diagram for $Pe = 30$ and $Ra \in \left[1.6\times10^4, 5.4 \times 10^4\right]$. In figure (a), the standard deviation of the $Nu(t)$ time series, $\sigma_{Nu}$, is plotted as function of $Ra$, and in (b) $\sigma_{Nu}$ is plotted as a function of $r = (Ra-Ra_1)/Ra_1$, where $Ra_1$ denotes the Rayleigh number at the bifurcation point and is $1.8 \times 10^4$ in this case. The circles are data points from simulations and the dashed line in (b) is the fit $\sigma_{Nu} = 0.47 \times r^{0.47 \pm 0.06}$.}
\label{fig:sigma2}
\end{centering}
\end{figure}

\begin{figure}
\begin{centering}
\includegraphics[trim = 0 0 0 0, width = \linewidth]{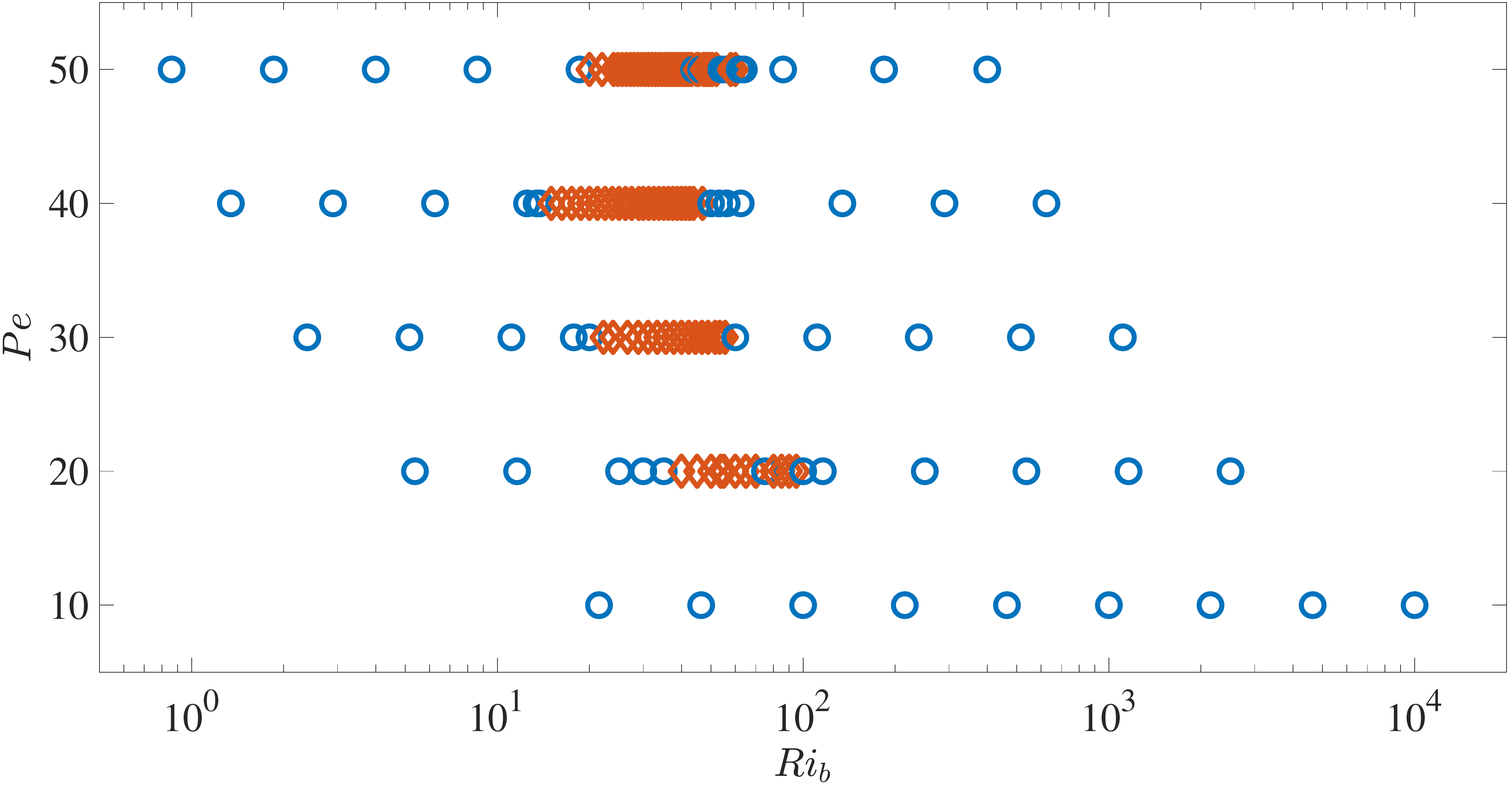} 
\caption{The $(Ri_b, Pe)$ phase diagram. Circles denote steady final states and diamonds denote periodic final states.}
\label{fig:phase}
\end{centering}
\end{figure}

The nature of this bifurcation can be established by studying how $\sigma_{Nu}$ changes with changing $r$. Here, $r = (Ra-Ra_1)/Ra_1$, where $Ra_1$ denotes the Rayleigh number at the bifurcation point. Figure \ref{fig:sigma1}(b) shows $\sigma_{Nu}$ as a function of $r$. Using a least-squares fit, one can determine that the increase in the amplitude close to the bifurcation point can be described using
\be
\sigma_{Nu} = 0.46 \times r^{0.52 \pm 0.12},
\ee
which is shown as the dashed line in figure \ref{fig:sigma1}(b). This is remarkably close to $\sigma_{Nu} \propto r^{0.5}$, which can be obtained from the solution of the Landau equation, which describes the time evolution of the amplitude of an unstable mode not far from the bifurcation point \citep{landau-fluid}. This, coupled with the fact that the bifurcation is from a steady to periodic state, leads us to conclude that this is a supercritical Hopf bifurcation. Although the transition from steady to periodic state is more gradual, the transition from periodic to steady state is relatively abrupt. Similar oscillatory states are observed for $Pe = 30, 40$ and $50$. In figure \ref{fig:sigma2}, the bifurcation diagram for $Pe=30$ is shown. A least squares fit to the data points close to the bifurcation point gives $\sigma_{Nu} = 0.47 \times r^{0.47 \pm 0.06}$, which is quantitatively similar to that obtained for $Pe = 20$.
The different windows of self-oscillations are shown in the $(Pe,Ri_b)$ phase diagram in figure \ref{fig:phase}. We should note that for $Pe = 20$ and $50$ there are multiple such windows.

\begin{figure}
\centering

\begin{subfigure}
\centering
\includegraphics[trim = 0 0 0 0, clip, width = \linewidth]{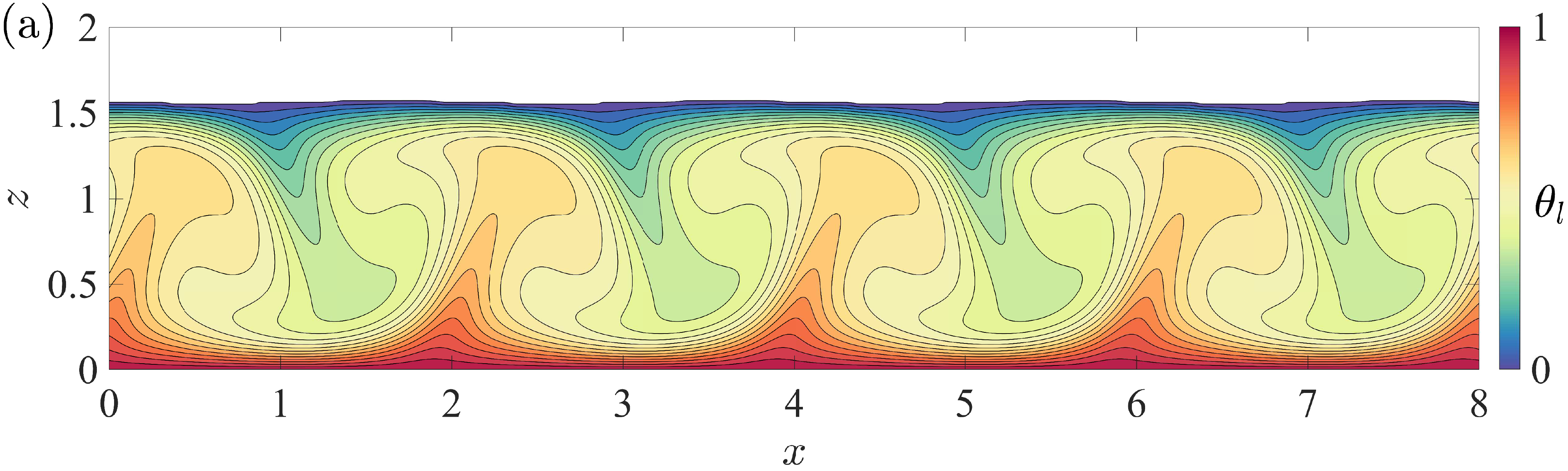}  
\end{subfigure}
    
\begin{subfigure}
\centering
\includegraphics[trim = 0 0 0 0, clip, width = \linewidth]{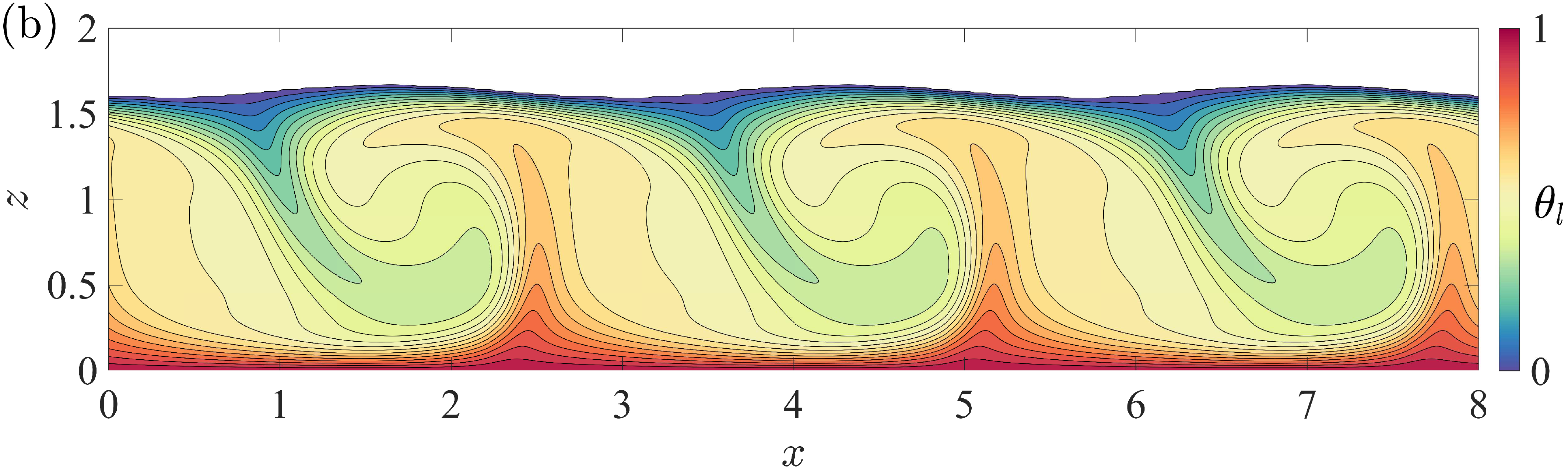} 
\end{subfigure}

\begin{subfigure}
\centering
\includegraphics[trim = 0 0 0 0, clip, width = \linewidth]{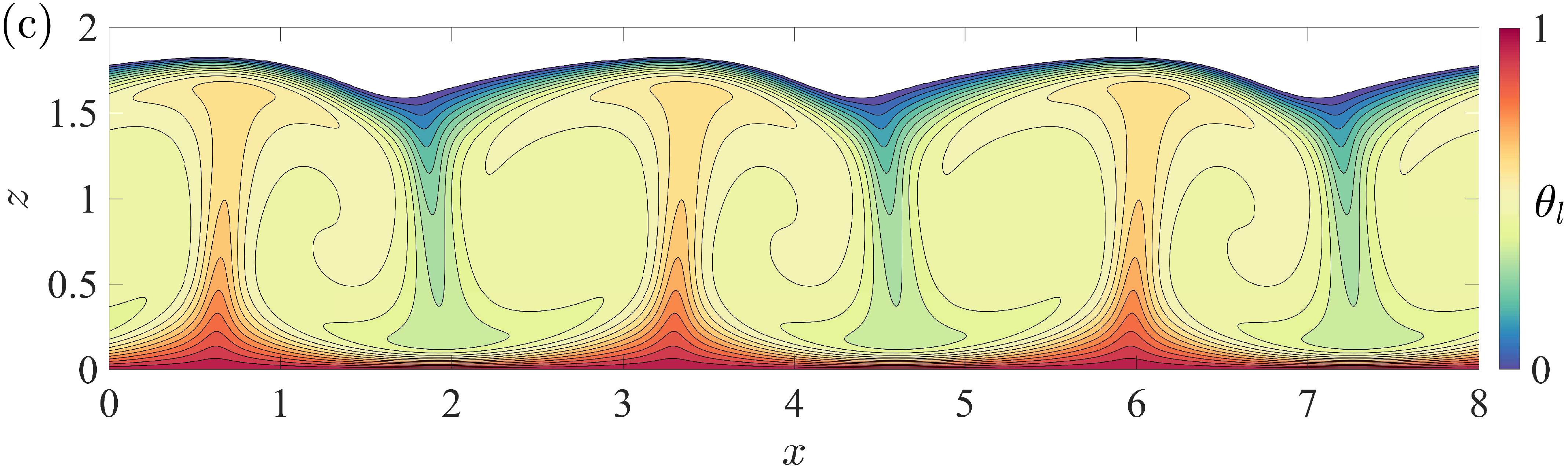} 
\end{subfigure}
                    
\caption{Temperature fields for $Pe = 20$ and (a) $Ra = 10^4$; (b) $Ra = 2.15\times 10^4$ and (c) $Ra = 4.64 \times 10^4$ in the stationary state. These values correspond to: (a) $Ri_b = 25$; (b) $Ri_b = 53.75$; and (c) $Ri_b = 116$. For $Ra = 10^4$ and $4.64 \times 10^4$, the plumes are frozen and the shear flow advects them; for $Ra = 2.15 \times 10^4$, the plumes oscillate about the vertical and are also advected by the shear flow. (Also see figure \ref{fig:phasetemp}.)}     
\label{fig:pattern}             
\end{figure}

\begin{figure}
\centering

\begin{subfigure}
\centering
\includegraphics[trim = 0 0 0 0, clip, width = \linewidth]{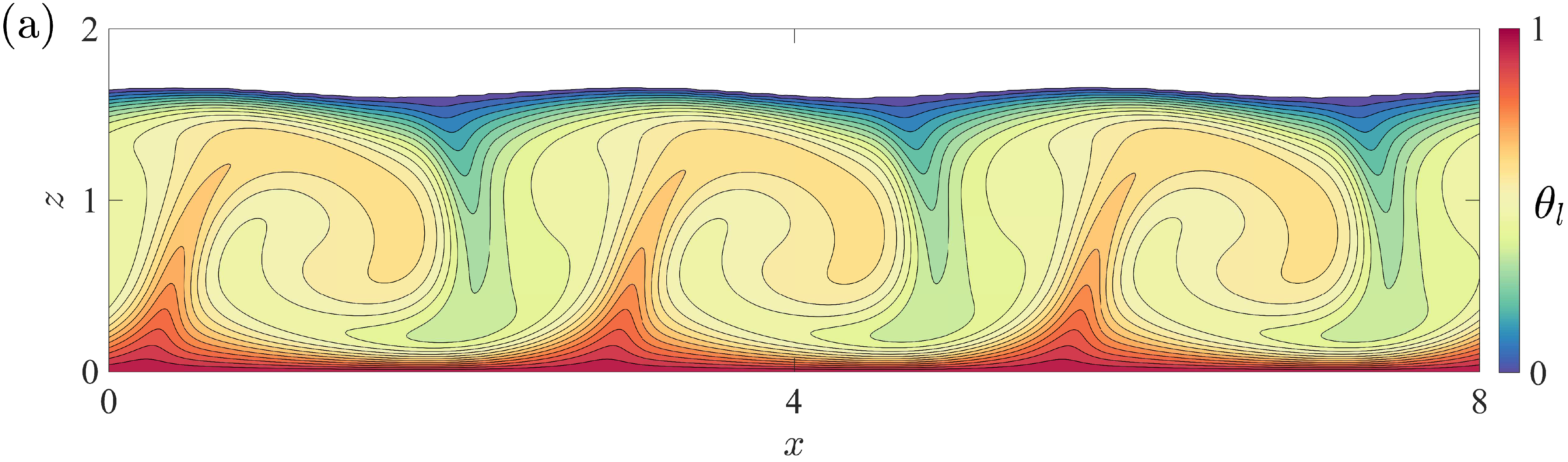}  
\end{subfigure}
    
\begin{subfigure}
\centering
\includegraphics[trim = 0 0 0 0, clip, width = \linewidth]{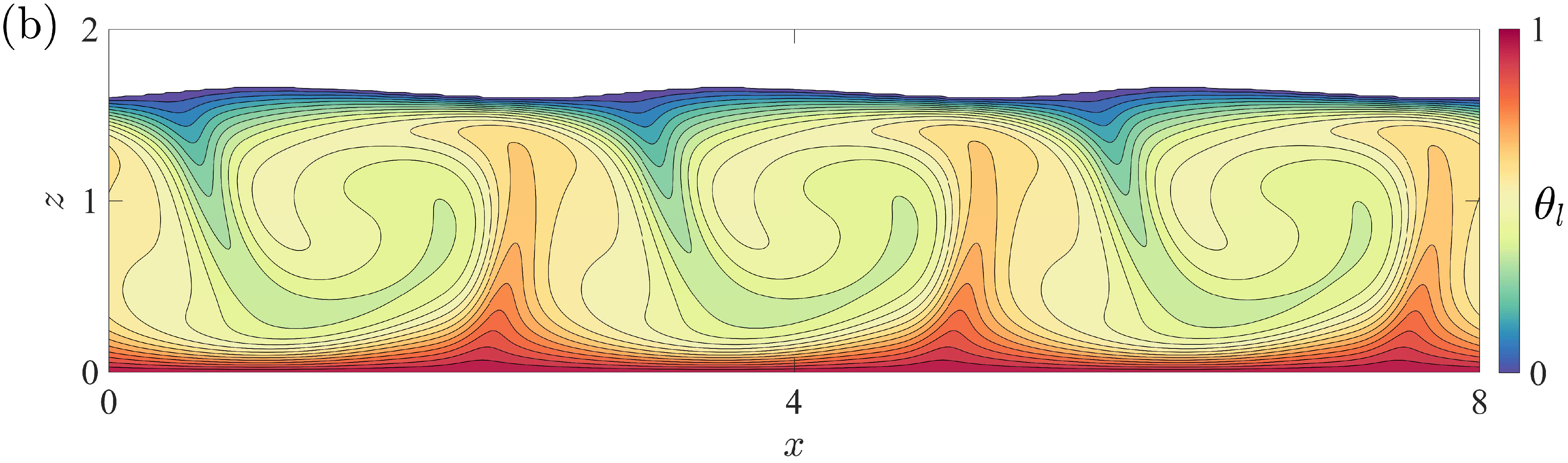} 
\end{subfigure}
                    
\caption{Snapshots of the temperature field for $Pe = 20$ and $Ra = 2.15 \times 10^4$ for: (a) $t = 1.87$ and (b) $t = 1.91$. These snapshots represent the temperature field at the maxima and minima of the time series in the inset of figure \ref{fig:pattern_time}.}     
\label{fig:phasetemp}             
\end{figure}

\begin{figure}
\centering
\begin{subfigure}
\centering
\includegraphics[trim = 0 0 0 0, clip, width = \linewidth]{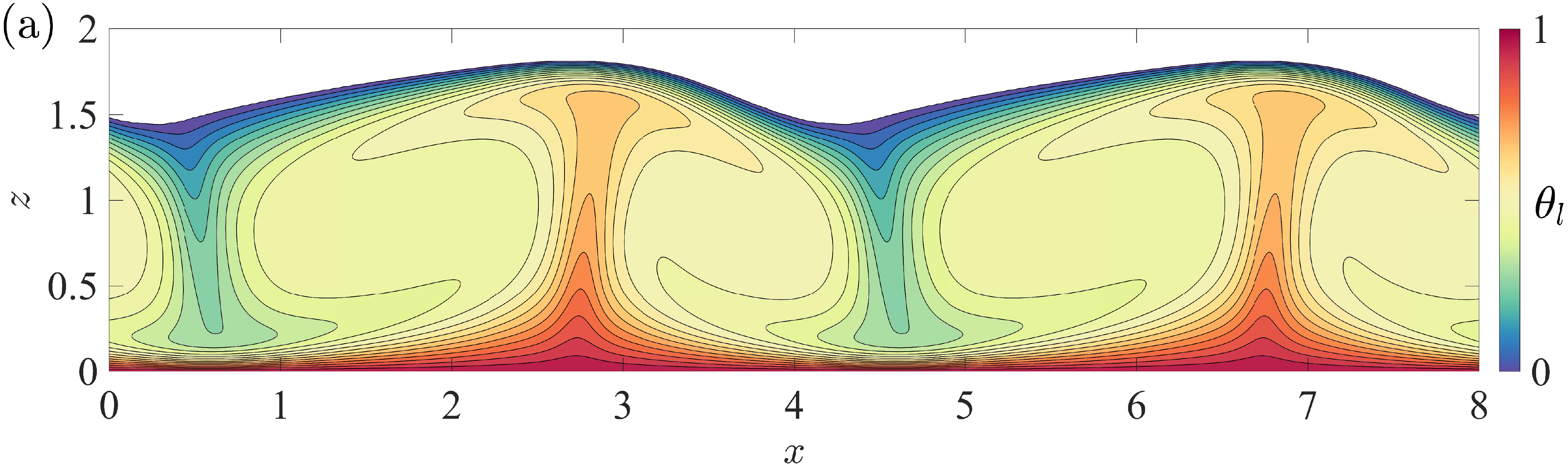}  
\end{subfigure}
    
\begin{subfigure}
\centering
\includegraphics[trim = 0 0 0 0, clip, width = \linewidth]{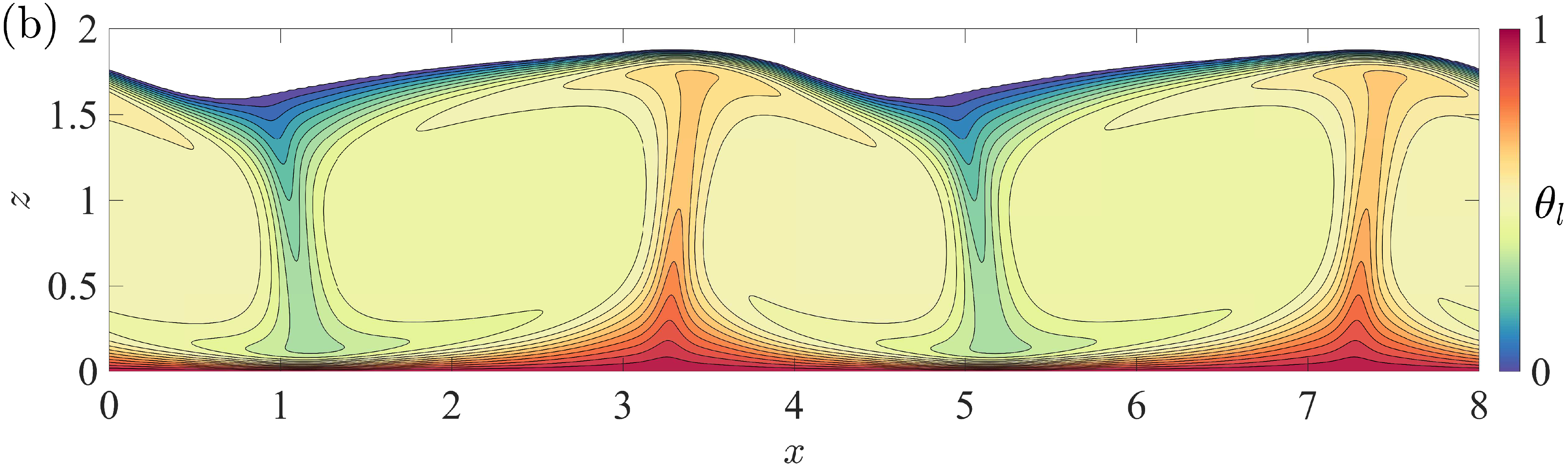} 
\end{subfigure}
                    
\caption{Snapshots of the temperature field for: (a) $Pe = 20$ and $Ra = 3 \times 10^4$ ($Ri_b = 75$) and (b) $Pe = 50$ and $Ra = 1.16 \times 10^5$ ($Ri_b = 46.4$) in the stationary state.}     
\label{fig:newpattern}             
\end{figure}

To understand the origin of this bifurcation, we study the temperature fields for the three cases of figure \ref{fig:pattern_time}, which are shown in figure \ref{fig:pattern}. We see that for $Ra = 10^4$ and $Ra = 4.64 \times 10^4$, the flow settles into a state with four and three pairs of convection cells, respectively. However, for $Ra = 2.15 \times 10^4$, the latter pattern is not stable, and results in the plumes oscillating about the vertical. These oscillations are due to the two competing spatial patterns  \citep[e.g.,][]{gollub1984} and can be seen in figures \ref{fig:phasetemp}(a) and \ref{fig:phasetemp}(b), which show the temperature fields at the maxima and minima of the $Nu(t)$ time series in the inset of figure \ref{fig:pattern_time}. This oscillatory behaviour can be discerned by observing the tilt of the cold plumes switch between leftwards and rightwards in \ref{fig:phasetemp}(a) and \ref{fig:phasetemp}(b), respectively. In the latter figure the plumes are more distorted, resulting in reduced vertical heat transport. We should also note here that such oscillatory behaviour is not observed when the fluid motions are purely convective.
For some of the stable states that occur between the periodic states in figure \ref{fig:phase}, we observe the stable flow pattern consists of only one pair of convection cells. This is shown in figures \ref{fig:newpattern}(a) and \ref{fig:newpattern}(b) for $Pe = 20$, $Ra = 3 \times 10^4$ and $Pe = 50$, $Ra = 1.16 \times 10^5$, respectively. 

This pattern competition can be understood by considering the principal effects of mean shear and buoyancy on the solid phase. For the range of $Pe$ studied here, the mean shear acts to inhibit vertical motions thereby melting less of the solid phase. This results in a relatively small change in the mean height of the liquid layer, thus preferring convection cells of smaller aspect ratio. However, buoyancy promotes vertical motions leading to more melting of the solid phase. This results in a larger change in the mean height of the liquid layer. Thus, in this case, the flow prefers convection cells of larger aspect ratio. The competition between these two effects is what leads to the observed pattern competition.

The multiple windows of self-oscillations for $Pe = 20$ and $50$ point to the possibility of existence of multiple solutions \footnote{This was suggested by one of the anonymous reviewers.}. However, for a given $Pe$ and $Ra$ (and other governing parameters) not all of these solutions might be stable. Hence the key question is: why does the system choose these specific solutions? This could be explored by using continuation methods to compute unstable solutions in order to understand the end result \citep[e.g.,][]{waleffe2015}. However, this is beyond the scope of the current work.

\subsubsection{Travelling interfacial waves}
One of the interesting results of \citet{Gilpin1980} is that under certain conditions a turbulent boundary layer flow gives rise to travelling waves at the phase boundary. In their experiments, the interfacial waves developed and propagated downstream over a period of 6 - 16 hours, depending on the Reynolds numbers and temperature boundary conditions. \citet{TW2019}, through their linear stability analysis of the Rayleigh-B\'enard-Couette flow over a phase boundary, showed that interfacial waves can be generated in the laminar regime close to $Ra = Ra_c$ for $Pe \in [0, 0.22]$. Hence, these waves can potentially be associated with the presence of a mean shear flow.

\begin{figure}
\begin{centering}
\includegraphics[trim = 0 0 0 0, width = \linewidth]{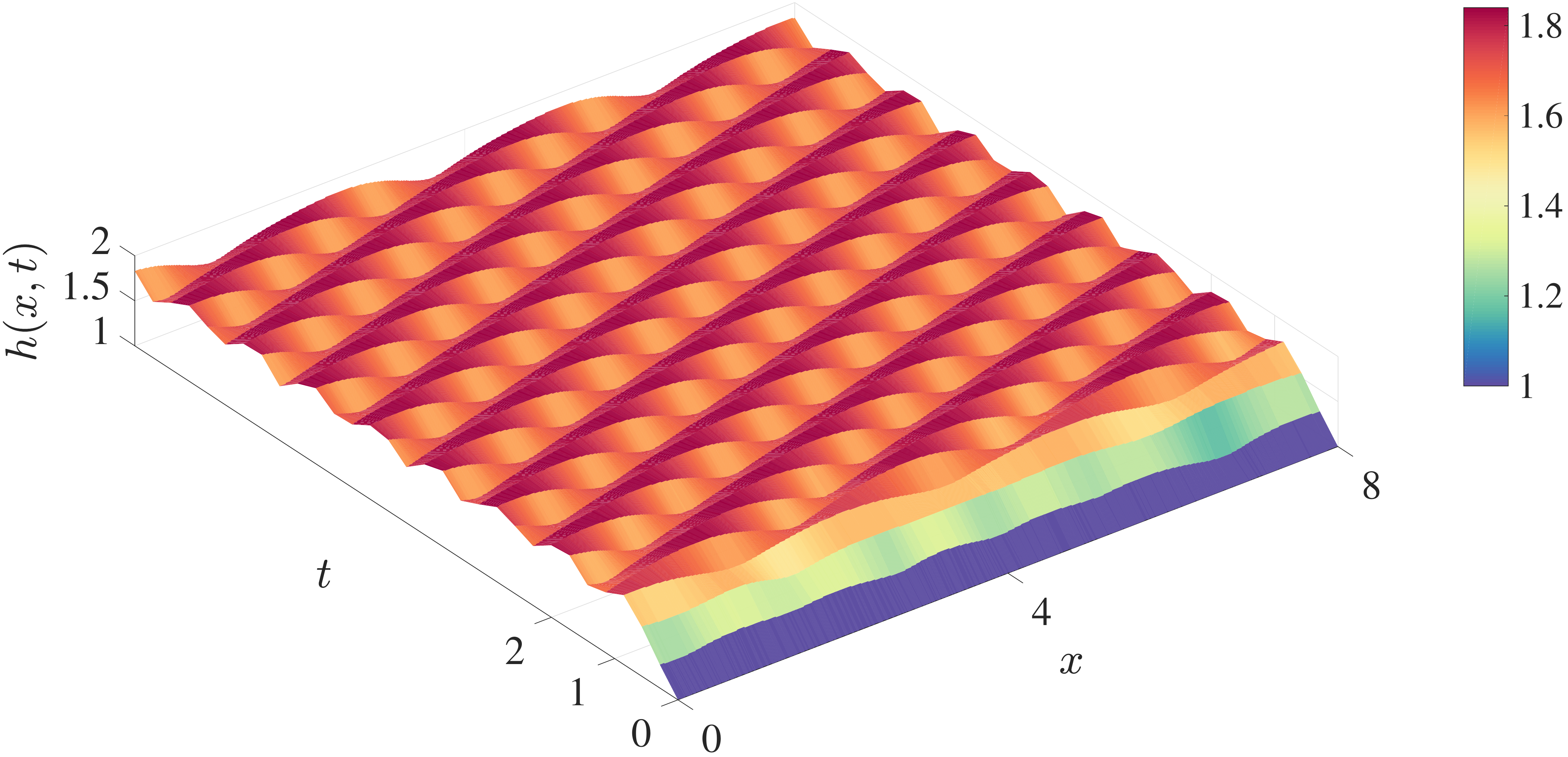} 
\caption{Spatio-temporal evolution of the interface for $Pe=20$ and $Ra = 4.64 \times 10^4$. The total duration of the simulation is $t = 8.58$. Any two neighbouring curves are separated by $\Delta t=0.28$.}
\label{fig:wave_xt}
\end{centering}
\end{figure}

In figure \ref{fig:wave_xt}, the spatio-temporal evolution of the phase boundary for $Pe=20$ and $Ra = 4.64 \times 10^4$ is shown. The total duration of the simulation is $t = 8.58$, and any two neighbouring curves are separated by $\Delta t = 0.28$. The presence of the interfacial wave is easily discerned by observing changes in the phase at a fixed $x$ location. The interfacial wave is propagating from left to right.

\begin{figure}
\centering

\begin{subfigure}
\centering
\includegraphics[trim = 0 0 0 0, clip, width = \linewidth]{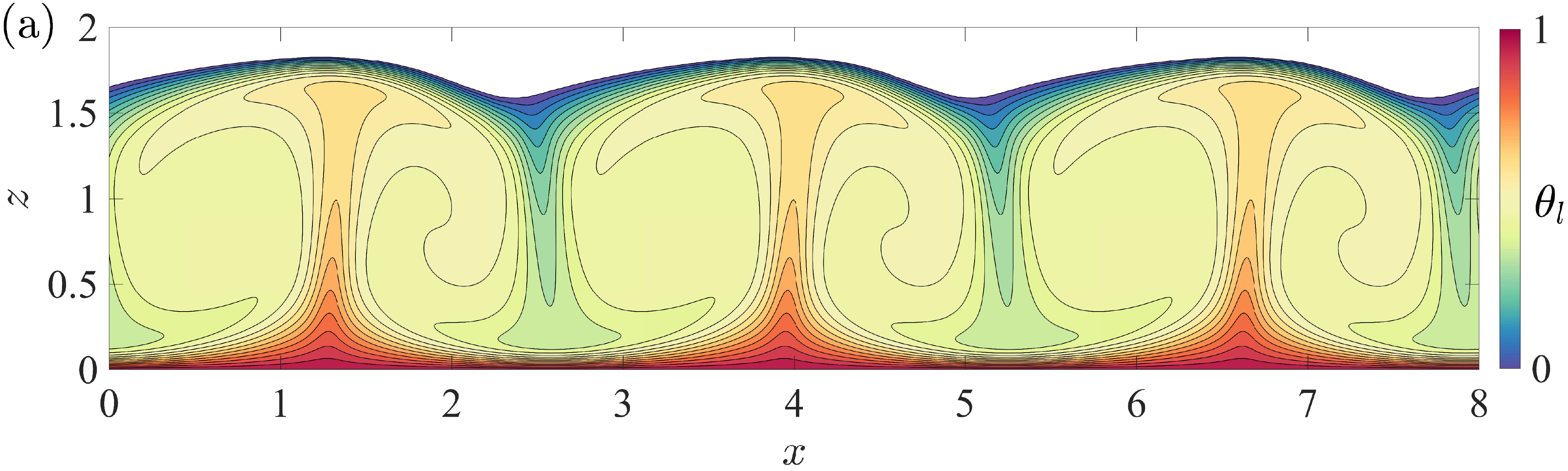}  
\end{subfigure}
    
\begin{subfigure}
\centering
\includegraphics[trim = 0 0 0 0, clip, width = \linewidth]{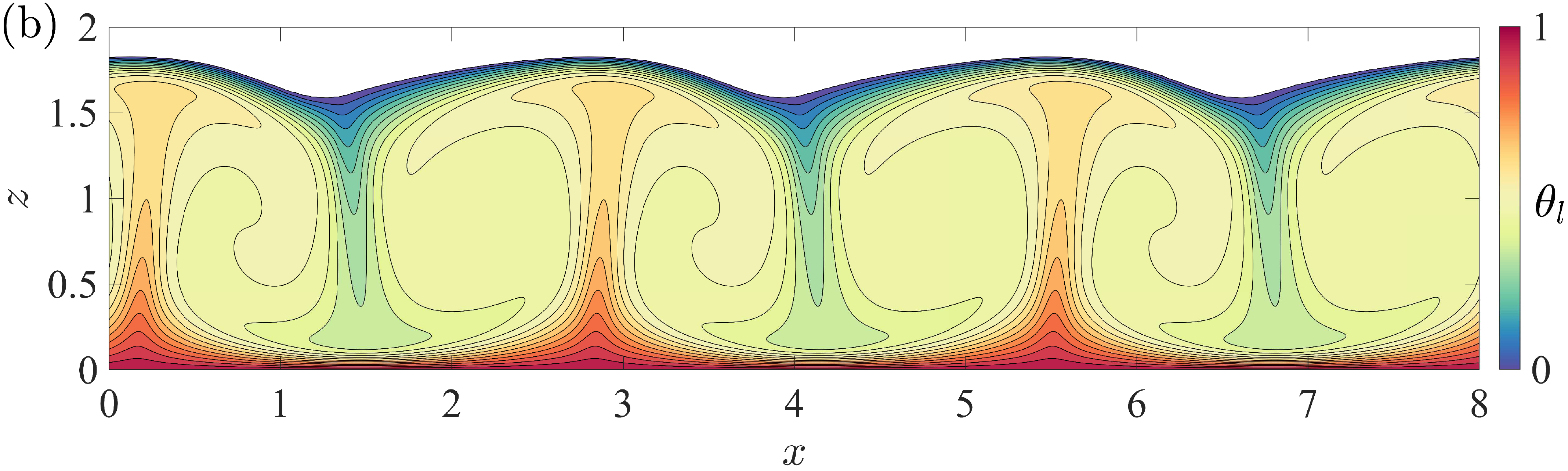} 
\end{subfigure}

\begin{subfigure}
\centering
\includegraphics[trim = 0 0 0 0, clip, width = \linewidth]{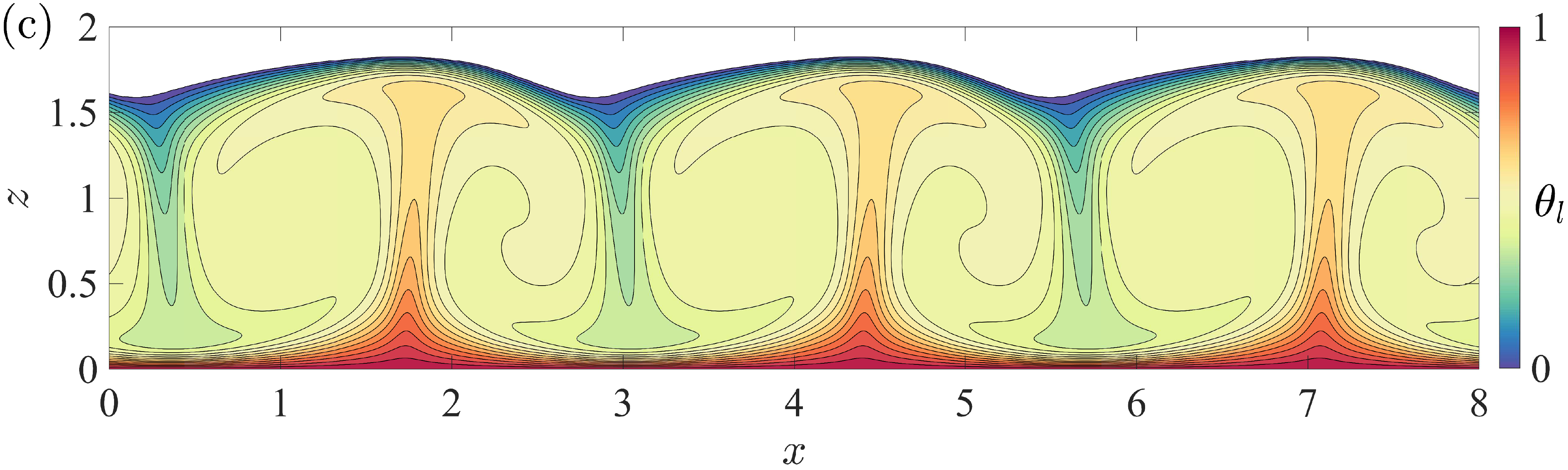} 
\end{subfigure}
                    
\caption{Travelling waves at the phase boundary for $Ra = 4.64 \times 10^4$ and $Pe = 20$. The temperature fields are for: (a) $t = 6.70$; (b) $t = 7.24$; and $t = 7.77$. Also see the movie in supplementary information.}     
\label{fig:wave}             
\end{figure}

To understand the mechanism of generation and propagation of this wave, we examine the evolution of the temperature field, which is shown in figure \ref{fig:wave}. Figures \ref{fig:wave}(a) -- \ref{fig:wave}(c) show snapshots of the temperature field for $Pe = 20$ and $Ra = 4.64 \times 10^4$ at three different times after the flow has reached a stationary state. Focussing on the hot plumes, one can see that they are advected along the domain by the Poiseuille flow. As they are advected, they locally melt some of the solid. The opposite is true for the cold plumes descending from the phase boundary: the solid grows locally as they are advected. This pattern of local growth and melting gives rise to the travelling wave that is seen in figure \ref{fig:wave_xt}. This also implies that the crests and troughs of the wave are locked in with the convection cells.

\begin{figure}
\begin{centering}
\includegraphics[trim = 0 0 0 0, width = \linewidth]{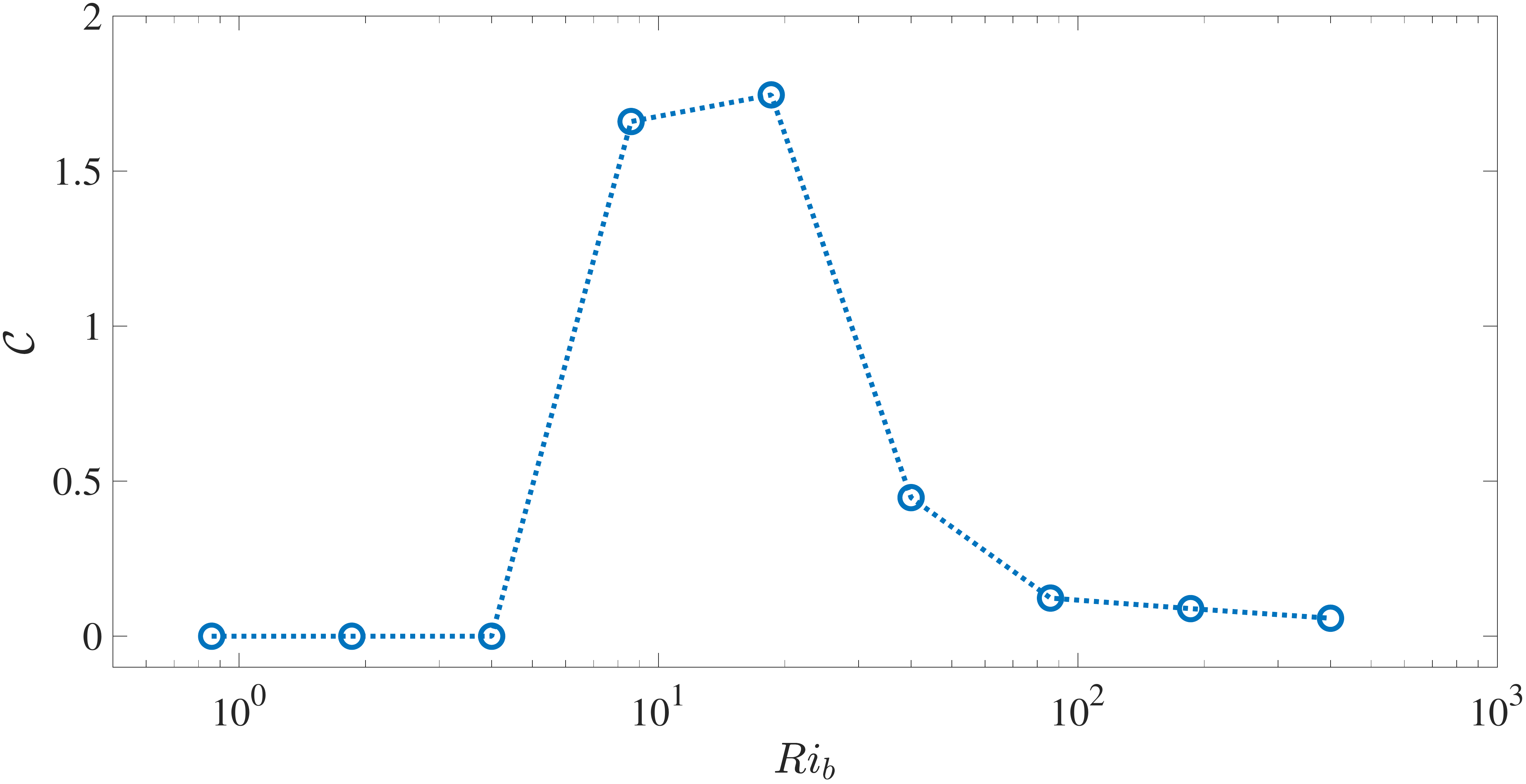} 
\caption{Phase speed of the interfacial waves for $Pe = 50$ as a function of $Ri_b$. The dimensional phase speed has been made dimensionless using $U_0$. For this reason, $\mathcal{C} \ge 1$ for certain values of $Ri_b$. However, as the flow and the phase boundary evolve, the maximum horizontal speed of the fluid increases. This is not reflected in the figure.}
\label{fig:wave_speed}
\end{centering}
\end{figure}

These waves can be further characterized by their non-dimensional phase speed $\mathcal{C}$, which is shown as a function of $Ri_b$ for $Pe = 50$ in figure \ref{fig:wave_speed}. Here, the dimensional phase speed has been non-dimensionalized using $U_0$. It is seen from figure \ref{fig:wave_speed} that for $Ri_b \ll 1$, $\mathcal{C} = 0$ and for $Ri_b \gg 1$, $C \ll 1$. This is because, for $Ri_b \ll 1$ the amplitude of the interfacial wave vanishes because no waves are formed; and, for $Ri_b \gg 1$ the mean shear flow is negligible. Hence, both mean shear and buoyancy are necessary to generate these travelling interfacial waves.

\subsubsection{Effects of large Stefan number on heat transport} \label{sec:stefan}
\begin{figure}
\begin{centering}
\includegraphics[trim = 0 0 0 0, width = \linewidth]{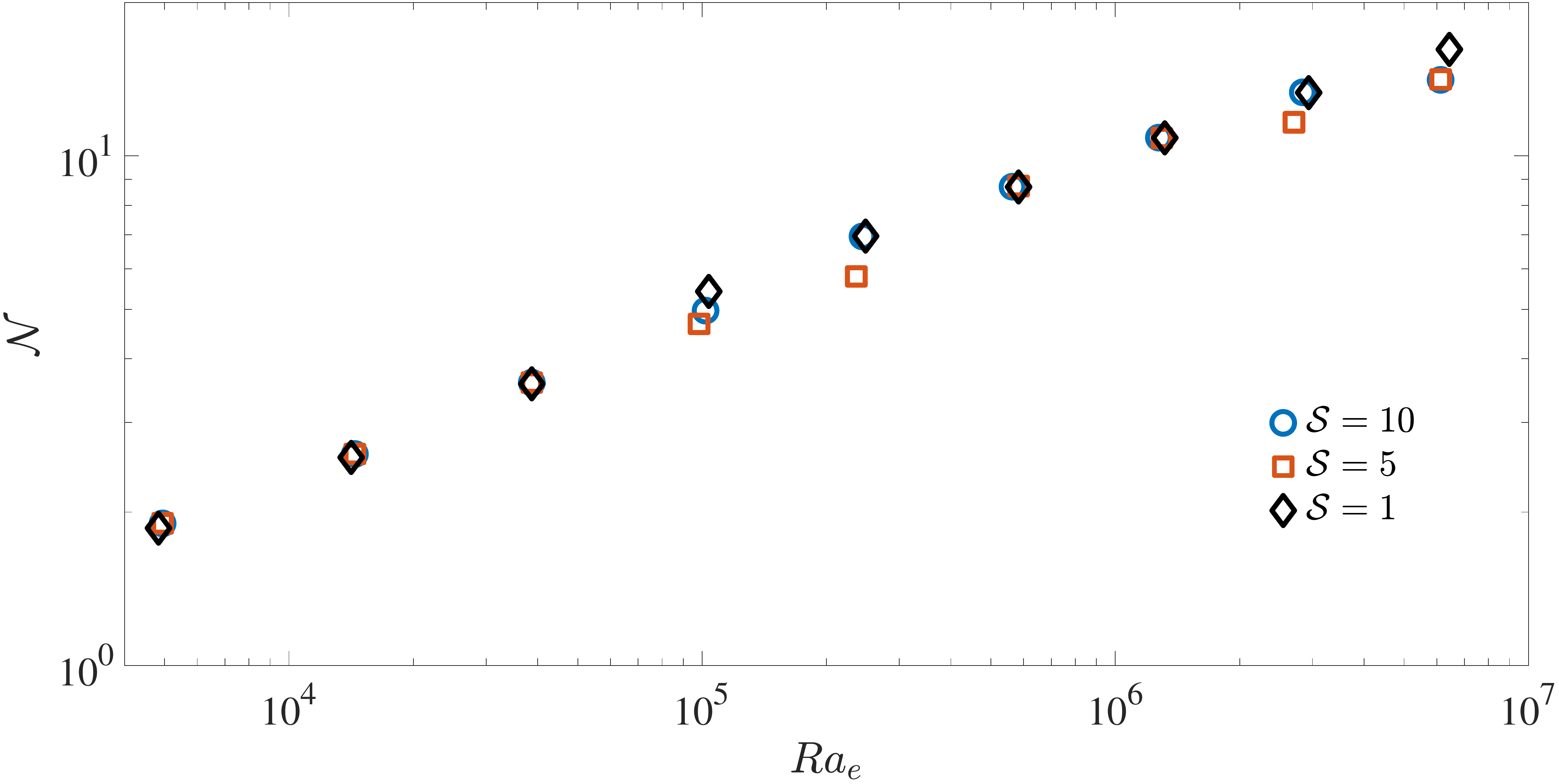} 
\caption{$\mathcal{N}$ vs. $Ra_e$ for $Pe = 10$ and the different values of $\mathcal{S}$.}
\label{fig:NuRa_Re10}
\end{centering}
\end{figure}

\begin{figure}
\begin{centering}
\includegraphics[trim = 0 0 0 0, width = \linewidth]{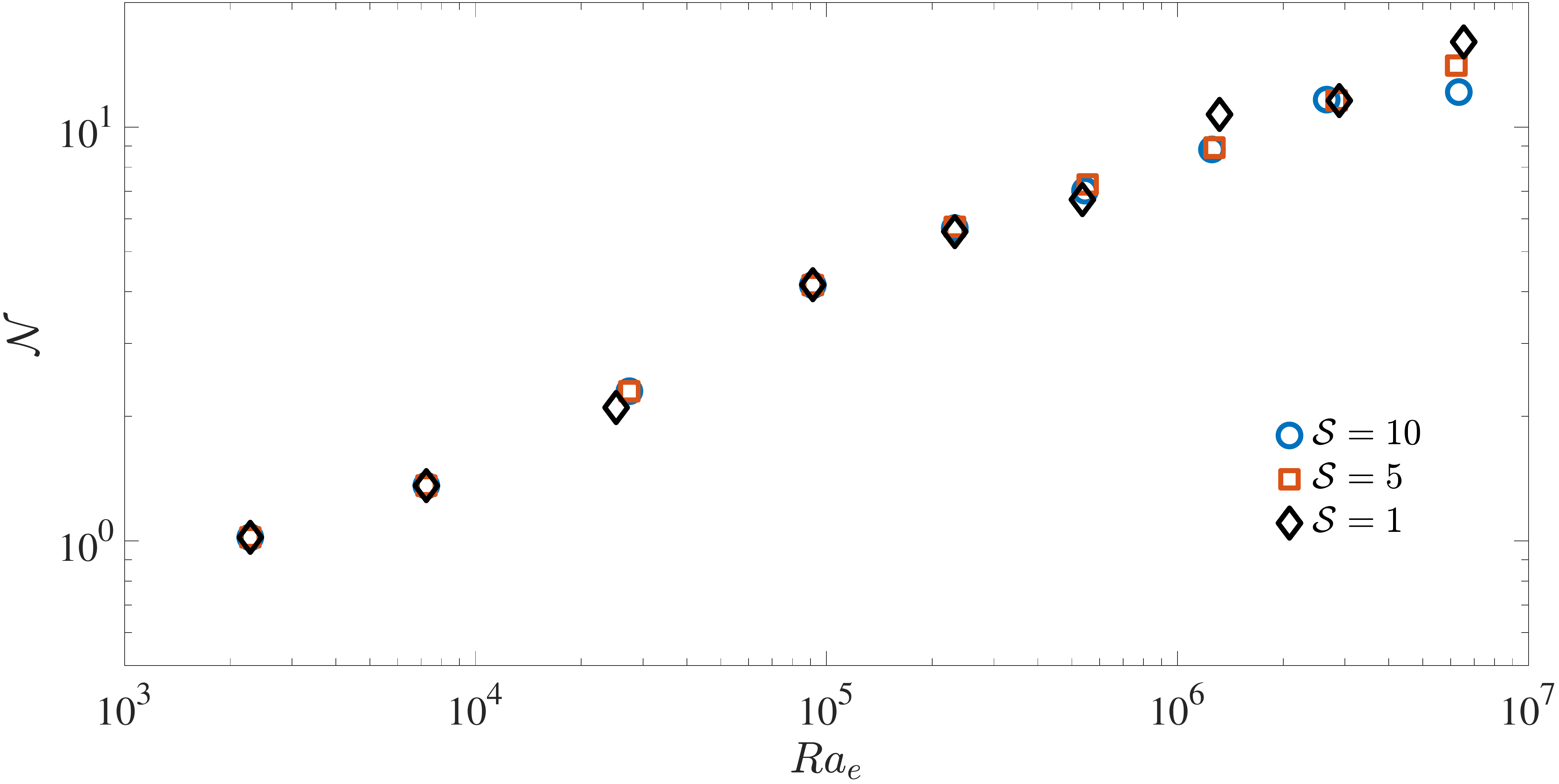} 
\caption{$\mathcal{N}$ vs. $Ra_e$ for $Pe = 50$ and the different values of $\mathcal{S}$.}
\label{fig:NuRa_Re50}
\end{centering}
\end{figure}

In many systems of interest, especially in geophysical settings \citep[e.g.,][]{MU71}, $\mathcal{S} \gg 1$. Hence, it is important to understand the effects of a large $\mathcal{S}$ on $\mathcal{N}$. In figures \ref{fig:NuRa_Re10} and \ref{fig:NuRa_Re50}, we show $\mathcal{N}$ as a function of $Ra_e$ for $Pe = 10$ and $50$, respectively, and three different values of $\mathcal{S}$. For both $Pe=10$ and $50$ the values of $\mathcal{N}$ for the different $\mathcal{S}$ are close to each other.
Hence, $\mathcal{S}$ does not seem to have a significant impact on the heat transport in this system. For a given $Pe$ and $Ra$, the small divergences that are seen in the values of $\mathcal{N}$ are due to variations in the mean depth of the liquid layer $h_m$. Convective motions tend to melt more of the solid phase and hence increase $h_m$, but mean shear and larger values of $\mathcal{S}$ tend to oppose it. The resulting $\mathcal{N}$ is due to a combination of these factors, and is clearly seen for the data points for $Ra_e \approx 6.5 \times 10^6$ in figure \ref{fig:NuRa_Re50}. This insensitivity is in qualitative agreement with the findings of \citet{esfahani2018}, who observed it in Rayleigh-B\'enard convection over a phase boundary.

\section{Conclusions}
We have systematically studied the effects of Rayleigh-B\'enard-Poiseuille flow on the evolution of a phase boundary in two dimensions using a combination of LBM and enthalpy method for the following range of control parameters: $Ra \in \left[2.15 \times 10^3, 10^6\right]$ and $Pe \in \left[0,50\right]$. The following are the main conclusions of our study:
\begin{enumerate}

\item The critical Rayleigh number and wavenumber for the onset of convection from our simulations were found to be in very good agreement with the results from the linear stability analysis of \citet{davis1984}.

\item For pure convection, the dependence of $\mathcal{N}$ on $Ra_e$ can be represented as a power law $\mathcal{N} = 0.2 \times Ra_e^{0.285 \pm 0.009}$ for $Ra_e \in \left[5.5 \times 10^3, 6.4 \times 10^6\right]$. The exponent $\beta = 0.285 \pm 0.009$ is in excellent agreement with the previous DNS studies of classical RBC \citep{doering2009, TSW2015_EPL}. The prefactor in the power law depends on the geometry \citep{TSW2015_EPL} and is larger than the prefactor for the classical RBC. Our $\mathcal{N}(Ra_e)$ data were also shown to be in good agreement with the results of \citet{purseed2020}.

\item Introduction of a Poiseuille flow was shown to considerably affect both the convective motions and the solid-liquid interface. The relative effects of mean shear and buoyancy were quantified using a bulk Richardson number, $Ri_b$. For $Ri_b = \mathcal{O}(1)$, the mean shear flow dominates and the transport of heat is only due to conduction. However, for $Ri_b \gg 1$ buoyancy has a dominating influence on the flow and on the evolution of the solid-liquid interface.

\item For moderate values of $Ri_b$, we observed travelling waves at the interface, in qualitative agreement with the experiments of \citet{Gilpin1980} and the linear stability analysis of \citet{TW2019}.

\item There are windows of self-oscillations for $Pe = 20, 30, 40$ and $50$ and $Ri_b \in \left[15,95\right]$, which are triggered by a pattern competition for convection cells of a certain aspect ratio. These oscillatory states were shown to occur through a supercritical Hopf bifurcation. However, such states were not observed for the case of purely convective flow.

\item We also explored the effects of larger values of $\mathcal{S}$ ($= 5$ and $10$) on the heat transport for $Pe = 10$ and $50$ and $Ra \in \left[2.15 \times 10^3, 10^6\right]$, and find that a large $\mathcal{S}$ does not have an appreciable impact on $\mathcal{N}$.

\end{enumerate}

The parameter phase space explored in this study was limited to laminar flows. The onset of unsteadiness and turbulence will have profound effects on the evolution of this system, and is a part of our future work.

\section*{Acknowledgements}
The author thanks A.J. Wells for helpful comments on an earlier draft of the manuscript and for suggesting figures \ref{fig:sigma1} and \ref{fig:sigma2}. The support of the University of Oxford and Yale University, through the facilities and staff of the Yale University Faculty of Arts and Sciences High Performance Computing Center, is gratefully acknowledged.

\section*{Declaration of interests}
The author reports no conflict of interest.

\appendix
\section{The enthalpy method} \label{sec:enthalpy}
In the enthalpy method, the total enthalpy is split into specific and latent heat contributions as:
\be
\mathcal{H}  = C_p \, T + L_s \, \phi,
\label{eqn:enthalpy}
\ee
where $\phi \in [0,1]$ is the liquid fraction of the concerned region. The enthalpies of pure liquid and solid phases at the melting point are $\mathcal{H_L} = C_p \, T_m + L_s$ and $\mathcal{H_S} = C_p \, T_m$, respectively. (The specific heats of the solid and liquid phases have been assumed to be the same.) The conservation equation for $\mathcal{H}$ when expressed in terms of $T$ using equation \ref{eqn:enthalpy} gives \citep{voller1987}
\be
\frac{\partial T}{\partial t} + \boldsymbol{u} \cdot \nabla T = \kappa \nabla^2 T - \frac{L_s}{C_p} \, \frac{\partial \phi}{\partial t}.
\label{eqn:enthalpy_T}
\ee
Equation \ref{eqn:enthalpy_T} combines the heat balance equation and the Stefan condition.

The following algorithm is used to calculate $T$ and $\phi$ numerically \citep{jiaung2001, huber2008}. When the temperature field is known at a time step $n$ and iteration $k$, the total enthalpy at a grid point $(i,j)$ is obtained by
\be
\mathcal{H}^{(n,k)}(i,j) = C_p \, T^{(n,k)}(i,j) + L_s \, \phi^{(n,k-1)}(i,j).
\ee
This is then used to determine the value of the $k^{\text{th}}$ iterate of $\phi$ using
\be
\phi^{(n,k)}(i,j) = \frac{\mathcal{H}^{(n,k)}(i,j) - \mathcal{H_S}}{\mathcal{H_L} - \mathcal{H_S}}.
\ee
If $\phi^{(n,k)}(i,j) < 0$ or $> 1$, then it is set to $0$ or $1$, respectively. This is then used to calculate $T^{(n,k+1)}(i,j)$. This process is repeated until converged values of $T$ and $\phi$, as determined by preset criteria, are obtained \citep{jiaung2001}.

In the LBM, the enthalpy method is implemented by introducing the source term in equation \ref{eqn:enthalpy_T} into the evolution equation for the temperature distribution functions \citep{jiaung2001, huber2008}. After the temperature field is calculated from the temperature distribution function, the steps outlined above are followed to update $\phi$. In our simulations, we find that using only one iteration provides results that are in good agreement with results obtained using phase-field method (see figure \ref{fig:NuRa_favier}). For this reason, we use only one iteration for all other calculations as well.

\bibliography{jfm_gilpin}

\begin{thebibliography}{50}%
\makeatletter
\providecommand \@ifxundefined [1]{%
 \@ifx{#1\undefined}
}%
\providecommand \@ifnum [1]{%
 \ifnum #1\expandafter \@firstoftwo
 \else \expandafter \@secondoftwo
 \fi
}%
\providecommand \@ifx [1]{%
 \ifx #1\expandafter \@firstoftwo
 \else \expandafter \@secondoftwo
 \fi
}%
\providecommand \natexlab [1]{#1}%
\providecommand \enquote  [1]{``#1''}%
\providecommand \bibnamefont  [1]{#1}%
\providecommand \bibfnamefont [1]{#1}%
\providecommand \citenamefont [1]{#1}%
\providecommand \href@noop [0]{\@secondoftwo}%
\providecommand \href [0]{\begingroup \@sanitize@url \@href}%
\providecommand \@href[1]{\@@startlink{#1}\@@href}%
\providecommand \@@href[1]{\endgroup#1\@@endlink}%
\providecommand \@sanitize@url [0]{\catcode `\\12\catcode `\$12\catcode
  `\&12\catcode `\#12\catcode `\^12\catcode `\_12\catcode `\%12\relax}%
\providecommand \@@startlink[1]{}%
\providecommand \@@endlink[0]{}%
\providecommand \url  [0]{\begingroup\@sanitize@url \@url }%
\providecommand \@url [1]{\endgroup\@href {#1}{\urlprefix }}%
\providecommand \urlprefix  [0]{URL }%
\providecommand \Eprint [0]{\href }%
\providecommand \doibase [0]{http://dx.doi.org/}%
\providecommand \selectlanguage [0]{\@gobble}%
\providecommand \bibinfo  [0]{\@secondoftwo}%
\providecommand \bibfield  [0]{\@secondoftwo}%
\providecommand \translation [1]{[#1]}%
\providecommand \BibitemOpen [0]{}%
\providecommand \bibitemStop [0]{}%
\providecommand \bibitemNoStop [0]{.\EOS\space}%
\providecommand \EOS [0]{\spacefactor3000\relax}%
\providecommand \BibitemShut  [1]{\csname bibitem#1\endcsname}%
\let\auto@bib@innerbib\@empty
\bibitem [{\citenamefont {Epstein}\ and\ \citenamefont
  {Cheung}(1983)}]{epstein1983}%
  \BibitemOpen
  \bibfield  {author} {\bibinfo {author} {\bibfnamefont {M.}~\bibnamefont
  {Epstein}}\ and\ \bibinfo {author} {\bibfnamefont {F.~B.}\ \bibnamefont
  {Cheung}},\ }\href@noop {} {\bibfield  {journal} {\bibinfo  {journal} {Ann.
  Rev. Fl. Mech.}\ }\textbf {\bibinfo {volume} {15}},\ \bibinfo {pages} {293}
  (\bibinfo {year} {1983})}\BibitemShut {NoStop}%
\bibitem [{\citenamefont {Glicksman}\ \emph {et~al.}(1986)\citenamefont
  {Glicksman}, \citenamefont {Coriell},\ and\ \citenamefont
  {McFadden}}]{glicksman1986}%
  \BibitemOpen
  \bibfield  {author} {\bibinfo {author} {\bibfnamefont {M.~E.}\ \bibnamefont
  {Glicksman}}, \bibinfo {author} {\bibfnamefont {S.~R.}\ \bibnamefont
  {Coriell}}, \ and\ \bibinfo {author} {\bibfnamefont {G.~B.}\ \bibnamefont
  {McFadden}},\ }\href@noop {} {\bibfield  {journal} {\bibinfo  {journal}
  {Annu. Rev. Fl. Mech.}\ }\textbf {\bibinfo {volume} {18}},\ \bibinfo {pages}
  {307} (\bibinfo {year} {1986})}\BibitemShut {NoStop}%
\bibitem [{\citenamefont {Huppert}(1986)}]{huppert1986}%
  \BibitemOpen
  \bibfield  {author} {\bibinfo {author} {\bibfnamefont {H.~E.}\ \bibnamefont
  {Huppert}},\ }\href@noop {} {\bibfield  {journal} {\bibinfo  {journal} {J.
  Fluid Mech.}\ }\textbf {\bibinfo {volume} {173}},\ \bibinfo {pages} {557}
  (\bibinfo {year} {1986})}\BibitemShut {NoStop}%
\bibitem [{\citenamefont {Worster}(2000)}]{worster2000}%
  \BibitemOpen
  \bibfield  {author} {\bibinfo {author} {\bibfnamefont {M.~G.}\ \bibnamefont
  {Worster}},\ }in\ \href@noop {} {\emph {\bibinfo {booktitle} {Perspectives in
  Fluid Dynamics --- a Collective Introduction to Current Research}}},\
  \bibinfo {editor} {edited by\ \bibinfo {editor} {\bibfnamefont
  {G.}~\bibnamefont {Batchelor}}, \bibinfo {editor} {\bibfnamefont
  {H.}~\bibnamefont {Moffatt}}, \ and\ \bibinfo {editor} {\bibfnamefont
  {M.}~\bibnamefont {Worster}}}\ (\bibinfo  {publisher} {Cambridge University
  Press},\ \bibinfo {year} {2000})\ pp.\ \bibinfo {pages} {393 --
  446}\BibitemShut {NoStop}%
\bibitem [{\citenamefont {Hewitt}(2020)}]{hewitt2020}%
  \BibitemOpen
  \bibfield  {author} {\bibinfo {author} {\bibfnamefont {I.~J.}\ \bibnamefont
  {Hewitt}},\ }\href@noop {} {\bibfield  {journal} {\bibinfo  {journal} {Annu.
  Rev. Fl. Mech.}\ }\textbf {\bibinfo {volume} {52}},\ \bibinfo {pages} {145}
  (\bibinfo {year} {2020})}\BibitemShut {NoStop}%
\bibitem [{\citenamefont {Davis}\ \emph {et~al.}(1984)\citenamefont {Davis},
  \citenamefont {M{\"u}ller},\ and\ \citenamefont {Dietsche}}]{davis1984}%
  \BibitemOpen
  \bibfield  {author} {\bibinfo {author} {\bibfnamefont {S.~H.}\ \bibnamefont
  {Davis}}, \bibinfo {author} {\bibfnamefont {U.}~\bibnamefont {M{\"u}ller}}, \
  and\ \bibinfo {author} {\bibfnamefont {C.}~\bibnamefont {Dietsche}},\
  }\href@noop {} {\bibfield  {journal} {\bibinfo  {journal} {J. Fluid Mech.}\
  }\textbf {\bibinfo {volume} {144}},\ \bibinfo {pages} {133} (\bibinfo {year}
  {1984})}\BibitemShut {NoStop}%
\bibitem [{\citenamefont {Dietsche}\ and\ \citenamefont
  {M{\"u}ller}(1985)}]{dietsche1985}%
  \BibitemOpen
  \bibfield  {author} {\bibinfo {author} {\bibfnamefont {C.}~\bibnamefont
  {Dietsche}}\ and\ \bibinfo {author} {\bibfnamefont {U.}~\bibnamefont
  {M{\"u}ller}},\ }\href@noop {} {\bibfield  {journal} {\bibinfo  {journal} {J.
  Fluid Mech.}\ }\textbf {\bibinfo {volume} {161}},\ \bibinfo {pages} {249}
  (\bibinfo {year} {1985})}\BibitemShut {NoStop}%
\bibitem [{\citenamefont {Wettlaufer}\ \emph {et~al.}(1997)\citenamefont
  {Wettlaufer}, \citenamefont {Worster},\ and\ \citenamefont
  {Huppert}}]{wettlaufer1997}%
  \BibitemOpen
  \bibfield  {author} {\bibinfo {author} {\bibfnamefont {J.~S.}\ \bibnamefont
  {Wettlaufer}}, \bibinfo {author} {\bibfnamefont {M.~G.}\ \bibnamefont
  {Worster}}, \ and\ \bibinfo {author} {\bibfnamefont {H.~E.}\ \bibnamefont
  {Huppert}},\ }\href@noop {} {\bibfield  {journal} {\bibinfo  {journal} {J.
  Fluid Mech.}\ }\textbf {\bibinfo {volume} {344}},\ \bibinfo {pages} {291}
  (\bibinfo {year} {1997})}\BibitemShut {NoStop}%
\bibitem [{\citenamefont {Worster}(1997)}]{worster1997}%
  \BibitemOpen
  \bibfield  {author} {\bibinfo {author} {\bibfnamefont {M.~G.}\ \bibnamefont
  {Worster}},\ }\href@noop {} {\bibfield  {journal} {\bibinfo  {journal} {Ann.
  Rev. Fl. Mech.}\ }\textbf {\bibinfo {volume} {29}},\ \bibinfo {pages} {91}
  (\bibinfo {year} {1997})}\BibitemShut {NoStop}%
\bibitem [{\citenamefont {Davies~Wykes}\ \emph {et~al.}(2018)\citenamefont
  {Davies~Wykes}, \citenamefont {Huang}, \citenamefont {Hajjar},\ and\
  \citenamefont {Ristroph}}]{wykes2018}%
  \BibitemOpen
  \bibfield  {author} {\bibinfo {author} {\bibfnamefont {M.~S.}\ \bibnamefont
  {Davies~Wykes}}, \bibinfo {author} {\bibfnamefont {J.~M.}\ \bibnamefont
  {Huang}}, \bibinfo {author} {\bibfnamefont {G.~A.}\ \bibnamefont {Hajjar}}, \
  and\ \bibinfo {author} {\bibfnamefont {L.}~\bibnamefont {Ristroph}},\
  }\href@noop {} {\bibfield  {journal} {\bibinfo  {journal} {Phys. Rev.
  Fluids}\ }\textbf {\bibinfo {volume} {3}},\ \bibinfo {pages} {043801}
  (\bibinfo {year} {2018})}\BibitemShut {NoStop}%
\bibitem [{\citenamefont {Delves}(1968)}]{delves1968}%
  \BibitemOpen
  \bibfield  {author} {\bibinfo {author} {\bibfnamefont {R.~T.}\ \bibnamefont
  {Delves}},\ }\href@noop {} {\bibfield  {journal} {\bibinfo  {journal} {J.
  Cryst. Growth}\ }\textbf {\bibinfo {volume} {3}},\ \bibinfo {pages} {562}
  (\bibinfo {year} {1968})}\BibitemShut {NoStop}%
\bibitem [{\citenamefont {Delves}(1971)}]{delves1971}%
  \BibitemOpen
  \bibfield  {author} {\bibinfo {author} {\bibfnamefont {R.~T.}\ \bibnamefont
  {Delves}},\ }\href@noop {} {\bibfield  {journal} {\bibinfo  {journal} {J.
  Cryst. Growth}\ }\textbf {\bibinfo {volume} {8}},\ \bibinfo {pages} {13}
  (\bibinfo {year} {1971})}\BibitemShut {NoStop}%
\bibitem [{\citenamefont {Gilpin}\ \emph {et~al.}(1980)\citenamefont {Gilpin},
  \citenamefont {Hirata},\ and\ \citenamefont {Cheng}}]{Gilpin1980}%
  \BibitemOpen
  \bibfield  {author} {\bibinfo {author} {\bibfnamefont {R.~R.}\ \bibnamefont
  {Gilpin}}, \bibinfo {author} {\bibfnamefont {T.}~\bibnamefont {Hirata}}, \
  and\ \bibinfo {author} {\bibfnamefont {K.~C.}\ \bibnamefont {Cheng}},\
  }\href@noop {} {\bibfield  {journal} {\bibinfo  {journal} {J. Fluid Mech.}\
  }\textbf {\bibinfo {volume} {99}},\ \bibinfo {pages} {619} (\bibinfo {year}
  {1980})}\BibitemShut {NoStop}%
\bibitem [{\citenamefont {Coriell}\ \emph {et~al.}(1984)\citenamefont
  {Coriell}, \citenamefont {McFadden}, \citenamefont {Boisvert},\ and\
  \citenamefont {Sekerka}}]{coriell1984}%
  \BibitemOpen
  \bibfield  {author} {\bibinfo {author} {\bibfnamefont {S.~R.}\ \bibnamefont
  {Coriell}}, \bibinfo {author} {\bibfnamefont {G.~B.}\ \bibnamefont
  {McFadden}}, \bibinfo {author} {\bibfnamefont {R.~F.}\ \bibnamefont
  {Boisvert}}, \ and\ \bibinfo {author} {\bibfnamefont {R.~F.}\ \bibnamefont
  {Sekerka}},\ }\href@noop {} {\bibfield  {journal} {\bibinfo  {journal} {J.
  Cryst. Growth}\ }\textbf {\bibinfo {volume} {69}},\ \bibinfo {pages} {15}
  (\bibinfo {year} {1984})}\BibitemShut {NoStop}%
\bibitem [{\citenamefont {Forth}\ and\ \citenamefont
  {Wheeler}(1989)}]{forth1989}%
  \BibitemOpen
  \bibfield  {author} {\bibinfo {author} {\bibfnamefont {S.~A.}\ \bibnamefont
  {Forth}}\ and\ \bibinfo {author} {\bibfnamefont {A.~A.}\ \bibnamefont
  {Wheeler}},\ }\href@noop {} {\bibfield  {journal} {\bibinfo  {journal} {J.
  Fluid Mech.}\ }\textbf {\bibinfo {volume} {202}},\ \bibinfo {pages} {339}
  (\bibinfo {year} {1989})}\BibitemShut {NoStop}%
\bibitem [{\citenamefont {Feltham}\ and\ \citenamefont
  {Worster}(1999)}]{feltham1999}%
  \BibitemOpen
  \bibfield  {author} {\bibinfo {author} {\bibfnamefont {D.~L.}\ \bibnamefont
  {Feltham}}\ and\ \bibinfo {author} {\bibfnamefont {M.~G.}\ \bibnamefont
  {Worster}},\ }\href@noop {} {\bibfield  {journal} {\bibinfo  {journal} {J.
  Fluid Mech.}\ }\textbf {\bibinfo {volume} {391}},\ \bibinfo {pages} {337}
  (\bibinfo {year} {1999})}\BibitemShut {NoStop}%
\bibitem [{\citenamefont {Neufeld}\ and\ \citenamefont
  {Wettlaufer}(2008{\natexlab{a}})}]{neufeld2008}%
  \BibitemOpen
  \bibfield  {author} {\bibinfo {author} {\bibfnamefont {J.~A.}\ \bibnamefont
  {Neufeld}}\ and\ \bibinfo {author} {\bibfnamefont {J.~S.}\ \bibnamefont
  {Wettlaufer}},\ }\href@noop {} {\bibfield  {journal} {\bibinfo  {journal} {J.
  Fluid Mech.}\ }\textbf {\bibinfo {volume} {612}},\ \bibinfo {pages} {363}
  (\bibinfo {year} {2008}{\natexlab{a}})}\BibitemShut {NoStop}%
\bibitem [{\citenamefont {Neufeld}\ and\ \citenamefont
  {Wettlaufer}(2008{\natexlab{b}})}]{neufeld2008shear}%
  \BibitemOpen
  \bibfield  {author} {\bibinfo {author} {\bibfnamefont {J.~A.}\ \bibnamefont
  {Neufeld}}\ and\ \bibinfo {author} {\bibfnamefont {J.~S.}\ \bibnamefont
  {Wettlaufer}},\ }\href@noop {} {\bibfield  {journal} {\bibinfo  {journal} {J.
  Fluid Mech.}\ }\textbf {\bibinfo {volume} {612}},\ \bibinfo {pages} {339}
  (\bibinfo {year} {2008}{\natexlab{b}})}\BibitemShut {NoStop}%
\bibitem [{\citenamefont {Ramudu}\ \emph {et~al.}(2016)\citenamefont {Ramudu},
  \citenamefont {Hirsh}, \citenamefont {Olson},\ and\ \citenamefont
  {Gnanadesikan}}]{ramudu2016}%
  \BibitemOpen
  \bibfield  {author} {\bibinfo {author} {\bibfnamefont {E.}~\bibnamefont
  {Ramudu}}, \bibinfo {author} {\bibfnamefont {B.~H.}\ \bibnamefont {Hirsh}},
  \bibinfo {author} {\bibfnamefont {P.}~\bibnamefont {Olson}}, \ and\ \bibinfo
  {author} {\bibfnamefont {A.}~\bibnamefont {Gnanadesikan}},\ }\href@noop {}
  {\bibfield  {journal} {\bibinfo  {journal} {J. Fluid Mech.}\ }\textbf
  {\bibinfo {volume} {798}},\ \bibinfo {pages} {572} (\bibinfo {year}
  {2016})}\BibitemShut {NoStop}%
\bibitem [{\citenamefont {Bushuk}\ \emph {et~al.}(2019)\citenamefont {Bushuk},
  \citenamefont {Holland}, \citenamefont {Stanton}, \citenamefont {Stern},\
  and\ \citenamefont {Gray}}]{bushuk2019}%
  \BibitemOpen
  \bibfield  {author} {\bibinfo {author} {\bibfnamefont {M.}~\bibnamefont
  {Bushuk}}, \bibinfo {author} {\bibfnamefont {D.~M.}\ \bibnamefont {Holland}},
  \bibinfo {author} {\bibfnamefont {T.~P.}\ \bibnamefont {Stanton}}, \bibinfo
  {author} {\bibfnamefont {A.}~\bibnamefont {Stern}}, \ and\ \bibinfo {author}
  {\bibfnamefont {C.}~\bibnamefont {Gray}},\ }\href@noop {} {\bibfield
  {journal} {\bibinfo  {journal} {J. Fluid Mech.}\ }\textbf {\bibinfo {volume}
  {873}},\ \bibinfo {pages} {942} (\bibinfo {year} {2019})}\BibitemShut
  {NoStop}%
\bibitem [{\citenamefont {Esfahani}\ \emph {et~al.}(2018)\citenamefont
  {Esfahani}, \citenamefont {Hirata}, \citenamefont {Berti},\ and\
  \citenamefont {Calzavarini}}]{esfahani2018}%
  \BibitemOpen
  \bibfield  {author} {\bibinfo {author} {\bibfnamefont {B.~R.}\ \bibnamefont
  {Esfahani}}, \bibinfo {author} {\bibfnamefont {S.~C.}\ \bibnamefont
  {Hirata}}, \bibinfo {author} {\bibfnamefont {S.}~\bibnamefont {Berti}}, \
  and\ \bibinfo {author} {\bibfnamefont {E.}~\bibnamefont {Calzavarini}},\
  }\href@noop {} {\bibfield  {journal} {\bibinfo  {journal} {Phys. Rev.
  Fluids}\ }\textbf {\bibinfo {volume} {3}},\ \bibinfo {pages} {053501}
  (\bibinfo {year} {2018})}\BibitemShut {NoStop}%
\bibitem [{\citenamefont {Favier}\ \emph {et~al.}(2019)\citenamefont {Favier},
  \citenamefont {Purseed},\ and\ \citenamefont {Duchemin}}]{favier2019}%
  \BibitemOpen
  \bibfield  {author} {\bibinfo {author} {\bibfnamefont {B.}~\bibnamefont
  {Favier}}, \bibinfo {author} {\bibfnamefont {J.}~\bibnamefont {Purseed}}, \
  and\ \bibinfo {author} {\bibfnamefont {L.}~\bibnamefont {Duchemin}},\
  }\href@noop {} {\bibfield  {journal} {\bibinfo  {journal} {J. Fluid Mech.}\
  }\textbf {\bibinfo {volume} {858}},\ \bibinfo {pages} {437} (\bibinfo {year}
  {2019})}\BibitemShut {NoStop}%
\bibitem [{\citenamefont {Toppaladoddi}\ \emph
  {et~al.}(2015{\natexlab{a}})\citenamefont {Toppaladoddi}, \citenamefont
  {Succi},\ and\ \citenamefont {Wettlaufer}}]{TSW2015_EPL}%
  \BibitemOpen
  \bibfield  {author} {\bibinfo {author} {\bibfnamefont {S.}~\bibnamefont
  {Toppaladoddi}}, \bibinfo {author} {\bibfnamefont {S.}~\bibnamefont {Succi}},
  \ and\ \bibinfo {author} {\bibfnamefont {J.~S.}\ \bibnamefont {Wettlaufer}},\
  }\href@noop {} {\bibfield  {journal} {\bibinfo  {journal} {EPL}\ }\textbf
  {\bibinfo {volume} {111}},\ \bibinfo {pages} {44005} (\bibinfo {year}
  {2015}{\natexlab{a}})}\BibitemShut {NoStop}%
\bibitem [{\citenamefont {Purseed}\ \emph {et~al.}(2020)\citenamefont
  {Purseed}, \citenamefont {Favier}, \citenamefont {Duchemin},\ and\
  \citenamefont {Hester}}]{purseed2020}%
  \BibitemOpen
  \bibfield  {author} {\bibinfo {author} {\bibfnamefont {J.}~\bibnamefont
  {Purseed}}, \bibinfo {author} {\bibfnamefont {B.}~\bibnamefont {Favier}},
  \bibinfo {author} {\bibfnamefont {L.}~\bibnamefont {Duchemin}}, \ and\
  \bibinfo {author} {\bibfnamefont {E.~W.}\ \bibnamefont {Hester}},\
  }\href@noop {} {\bibfield  {journal} {\bibinfo  {journal} {Phys. Rev.
  Fluids}\ }\textbf {\bibinfo {volume} {5}},\ \bibinfo {pages} {023501}
  (\bibinfo {year} {2020})}\BibitemShut {NoStop}%
\bibitem [{\citenamefont {Toppaladoddi}\ and\ \citenamefont
  {Wettlaufer}(2019)}]{TW2019}%
  \BibitemOpen
  \bibfield  {author} {\bibinfo {author} {\bibfnamefont {S.}~\bibnamefont
  {Toppaladoddi}}\ and\ \bibinfo {author} {\bibfnamefont {J.~S.}\ \bibnamefont
  {Wettlaufer}},\ }\href@noop {} {\bibfield  {journal} {\bibinfo  {journal} {J.
  Fluid Mech.}\ }\textbf {\bibinfo {volume} {868}},\ \bibinfo {pages} {648}
  (\bibinfo {year} {2019})}\BibitemShut {NoStop}%
\bibitem [{\citenamefont {Hirata}\ \emph
  {et~al.}(1979{\natexlab{a}})\citenamefont {Hirata}, \citenamefont {Gilpin},\
  and\ \citenamefont {Cheng}}]{hirata1979a}%
  \BibitemOpen
  \bibfield  {author} {\bibinfo {author} {\bibfnamefont {T.}~\bibnamefont
  {Hirata}}, \bibinfo {author} {\bibfnamefont {R.~R.}\ \bibnamefont {Gilpin}},
  \ and\ \bibinfo {author} {\bibfnamefont {K.~C.}\ \bibnamefont {Cheng}},\
  }\href@noop {} {\bibfield  {journal} {\bibinfo  {journal} {Int. J. Heat Mass
  Transfer}\ }\textbf {\bibinfo {volume} {22}},\ \bibinfo {pages} {1435}
  (\bibinfo {year} {1979}{\natexlab{a}})}\BibitemShut {NoStop}%
\bibitem [{\citenamefont {Hirata}\ \emph
  {et~al.}(1979{\natexlab{b}})\citenamefont {Hirata}, \citenamefont {Gilpin},
  \citenamefont {Cheng},\ and\ \citenamefont {Gates}}]{hirata1979}%
  \BibitemOpen
  \bibfield  {author} {\bibinfo {author} {\bibfnamefont {T.}~\bibnamefont
  {Hirata}}, \bibinfo {author} {\bibfnamefont {R.~R.}\ \bibnamefont {Gilpin}},
  \bibinfo {author} {\bibfnamefont {K.~C.}\ \bibnamefont {Cheng}}, \ and\
  \bibinfo {author} {\bibfnamefont {E.~M.}\ \bibnamefont {Gates}},\ }\href@noop
  {} {\bibfield  {journal} {\bibinfo  {journal} {Int. J. Heat Mass Transfer}\
  }\textbf {\bibinfo {volume} {22}},\ \bibinfo {pages} {1425} (\bibinfo {year}
  {1979}{\natexlab{b}})}\BibitemShut {NoStop}%
\bibitem [{\citenamefont {Monin}\ and\ \citenamefont
  {Yaglom}(1971)}]{Monin1971}%
  \BibitemOpen
  \bibfield  {author} {\bibinfo {author} {\bibfnamefont {A.}~\bibnamefont
  {Monin}}\ and\ \bibinfo {author} {\bibfnamefont {A.}~\bibnamefont {Yaglom}},\
  }\href@noop {} {\emph {\bibinfo {title} {Statistical fluid mechanics:
  Mechanics of turbulence volume 1}}}\ (\bibinfo  {publisher} {Dover
  Publications},\ \bibinfo {year} {1971})\BibitemShut {NoStop}%
\bibitem [{\citenamefont {Couston}\ \emph {et~al.}(2020)\citenamefont
  {Couston}, \citenamefont {Hester}, \citenamefont {Favier}, \citenamefont
  {Taylor}, \citenamefont {Holland},\ and\ \citenamefont
  {Jenkins}}]{couston2020}%
  \BibitemOpen
  \bibfield  {author} {\bibinfo {author} {\bibfnamefont {L.-A.}\ \bibnamefont
  {Couston}}, \bibinfo {author} {\bibfnamefont {E.}~\bibnamefont {Hester}},
  \bibinfo {author} {\bibfnamefont {B.}~\bibnamefont {Favier}}, \bibinfo
  {author} {\bibfnamefont {J.~R.}\ \bibnamefont {Taylor}}, \bibinfo {author}
  {\bibfnamefont {P.~R.}\ \bibnamefont {Holland}}, \ and\ \bibinfo {author}
  {\bibfnamefont {A.}~\bibnamefont {Jenkins}},\ }\href@noop {} {\bibfield
  {journal} {\bibinfo  {journal} {arXiv preprint arXiv:2004.09879}\ } (\bibinfo
  {year} {2020})}\BibitemShut {NoStop}%
\bibitem [{Note1()}]{Note1}%
  \BibitemOpen
  \bibinfo {note} {Except for velocity, we follow \protect \citet {TW2019} in
  choosing the different scales for non-dimensionalization.}\BibitemShut
  {Stop}%
\bibitem [{\citenamefont {Benzi}\ \emph {et~al.}(1992)\citenamefont {Benzi},
  \citenamefont {Succi},\ and\ \citenamefont {Vergassola}}]{benzi1992}%
  \BibitemOpen
  \bibfield  {author} {\bibinfo {author} {\bibfnamefont {R.}~\bibnamefont
  {Benzi}}, \bibinfo {author} {\bibfnamefont {S.}~\bibnamefont {Succi}}, \ and\
  \bibinfo {author} {\bibfnamefont {M.}~\bibnamefont {Vergassola}},\
  }\href@noop {} {\bibfield  {journal} {\bibinfo  {journal} {Phys. Rep.}\
  }\textbf {\bibinfo {volume} {222}},\ \bibinfo {pages} {145} (\bibinfo {year}
  {1992})}\BibitemShut {NoStop}%
\bibitem [{\citenamefont {Chen}\ and\ \citenamefont {Doolen}(1998)}]{chen1998}%
  \BibitemOpen
  \bibfield  {author} {\bibinfo {author} {\bibfnamefont {S.}~\bibnamefont
  {Chen}}\ and\ \bibinfo {author} {\bibfnamefont {G.~D.}\ \bibnamefont
  {Doolen}},\ }\href@noop {} {\bibfield  {journal} {\bibinfo  {journal} {Ann.
  Rev. Fluid Mech.}\ }\textbf {\bibinfo {volume} {30}},\ \bibinfo {pages} {329}
  (\bibinfo {year} {1998})}\BibitemShut {NoStop}%
\bibitem [{\citenamefont {Voller}\ \emph {et~al.}(1987)\citenamefont {Voller},
  \citenamefont {Cross},\ and\ \citenamefont {Markatos}}]{voller1987}%
  \BibitemOpen
  \bibfield  {author} {\bibinfo {author} {\bibfnamefont {V.~R.}\ \bibnamefont
  {Voller}}, \bibinfo {author} {\bibfnamefont {M.}~\bibnamefont {Cross}}, \
  and\ \bibinfo {author} {\bibfnamefont {N.~C.}\ \bibnamefont {Markatos}},\
  }\href@noop {} {\bibfield  {journal} {\bibinfo  {journal} {Int. J. Numer.
  Meth. Eng.}\ }\textbf {\bibinfo {volume} {24}},\ \bibinfo {pages} {271}
  (\bibinfo {year} {1987})}\BibitemShut {NoStop}%
\bibitem [{\citenamefont {Voller}\ and\ \citenamefont
  {Cross}(1981)}]{voller1981}%
  \BibitemOpen
  \bibfield  {author} {\bibinfo {author} {\bibfnamefont {V.}~\bibnamefont
  {Voller}}\ and\ \bibinfo {author} {\bibfnamefont {M.}~\bibnamefont {Cross}},\
  }\href@noop {} {\bibfield  {journal} {\bibinfo  {journal} {Int. J. Heat Mass
  Transfer}\ }\textbf {\bibinfo {volume} {24}},\ \bibinfo {pages} {545}
  (\bibinfo {year} {1981})}\BibitemShut {NoStop}%
\bibitem [{\citenamefont {Jiaung}\ \emph {et~al.}(2001)\citenamefont {Jiaung},
  \citenamefont {Ho},\ and\ \citenamefont {Kuo}}]{jiaung2001}%
  \BibitemOpen
  \bibfield  {author} {\bibinfo {author} {\bibfnamefont {W.-S.}\ \bibnamefont
  {Jiaung}}, \bibinfo {author} {\bibfnamefont {J.-R.}\ \bibnamefont {Ho}}, \
  and\ \bibinfo {author} {\bibfnamefont {C.-P.}\ \bibnamefont {Kuo}},\
  }\href@noop {} {\bibfield  {journal} {\bibinfo  {journal} {Numer. Heat
  Transf.: Part B}\ }\textbf {\bibinfo {volume} {39}},\ \bibinfo {pages} {167}
  (\bibinfo {year} {2001})}\BibitemShut {NoStop}%
\bibitem [{\citenamefont {Huber}\ \emph {et~al.}(2008)\citenamefont {Huber},
  \citenamefont {Parmigiani}, \citenamefont {Chopard}, \citenamefont {Manga},\
  and\ \citenamefont {Bachmann}}]{huber2008}%
  \BibitemOpen
  \bibfield  {author} {\bibinfo {author} {\bibfnamefont {C.}~\bibnamefont
  {Huber}}, \bibinfo {author} {\bibfnamefont {A.}~\bibnamefont {Parmigiani}},
  \bibinfo {author} {\bibfnamefont {B.}~\bibnamefont {Chopard}}, \bibinfo
  {author} {\bibfnamefont {M.}~\bibnamefont {Manga}}, \ and\ \bibinfo {author}
  {\bibfnamefont {O.}~\bibnamefont {Bachmann}},\ }\href@noop {} {\bibfield
  {journal} {\bibinfo  {journal} {Int. J. Heat Fluid Flow}\ }\textbf {\bibinfo
  {volume} {29}},\ \bibinfo {pages} {1469} (\bibinfo {year}
  {2008})}\BibitemShut {NoStop}%
\bibitem [{\citenamefont {Succi}(2001)}]{succi2001}%
  \BibitemOpen
  \bibfield  {author} {\bibinfo {author} {\bibfnamefont {S.}~\bibnamefont
  {Succi}},\ }\href@noop {} {\emph {\bibinfo {title} {The Lattice-Boltzmann
  Equation}}}\ (\bibinfo  {publisher} {Oxford University Press},\ \bibinfo
  {year} {2001})\BibitemShut {NoStop}%
\bibitem [{\citenamefont {Latt}(2007)}]{latt}%
  \BibitemOpen
  \bibfield  {author} {\bibinfo {author} {\bibfnamefont {J.}~\bibnamefont
  {Latt}},\ }\emph {\bibinfo {title} {Hydrodynamic limit of lattice {B}oltzmann
  equations}},\ \href@noop {} {Ph.D. thesis},\ \bibinfo  {school} {Universit\'e
  de Gen\`eve} (\bibinfo {year} {2007})\BibitemShut {NoStop}%
\bibitem [{\citenamefont {Toppaladoddi}(2017)}]{Toppaladoddi_PhD}%
  \BibitemOpen
  \bibfield  {author} {\bibinfo {author} {\bibfnamefont {S.}~\bibnamefont
  {Toppaladoddi}},\ }\emph {\bibinfo {title} {The staistical physics, fluid
  mechanics, and the climatology of {A}rctic sea ice}},\ \href@noop {} {Ph.D.
  thesis},\ \bibinfo  {school} {Yale University} (\bibinfo {year}
  {2017})\BibitemShut {NoStop}%
\bibitem [{\citenamefont {Toppaladoddi}\ \emph
  {et~al.}(2015{\natexlab{b}})\citenamefont {Toppaladoddi}, \citenamefont
  {Succi},\ and\ \citenamefont {Wettlaufer}}]{TSW2015_iutam}%
  \BibitemOpen
  \bibfield  {author} {\bibinfo {author} {\bibfnamefont {S.}~\bibnamefont
  {Toppaladoddi}}, \bibinfo {author} {\bibfnamefont {S.}~\bibnamefont {Succi}},
  \ and\ \bibinfo {author} {\bibfnamefont {J.~S.}\ \bibnamefont {Wettlaufer}},\
  }\href@noop {} {\bibfield  {journal} {\bibinfo  {journal} {Procedia IUTAM}\
  }\textbf {\bibinfo {volume} {15}},\ \bibinfo {pages} {34} (\bibinfo {year}
  {2015}{\natexlab{b}})}\BibitemShut {NoStop}%
\bibitem [{\citenamefont {Chandrasekhar}(2013)}]{chandra2013}%
  \BibitemOpen
  \bibfield  {author} {\bibinfo {author} {\bibfnamefont {S.}~\bibnamefont
  {Chandrasekhar}},\ }\href@noop {} {\emph {\bibinfo {title} {Hydrodynamic and
  {H}ydromagnetic {S}tability}}}\ (\bibinfo  {publisher} {Dover Publications},\
  \bibinfo {year} {2013})\BibitemShut {NoStop}%
\bibitem [{\citenamefont {Johnston}\ and\ \citenamefont
  {Doering}(2009)}]{doering2009}%
  \BibitemOpen
  \bibfield  {author} {\bibinfo {author} {\bibfnamefont {H.}~\bibnamefont
  {Johnston}}\ and\ \bibinfo {author} {\bibfnamefont {C.~R.}\ \bibnamefont
  {Doering}},\ }\href@noop {} {\bibfield  {journal} {\bibinfo  {journal} {Phys.
  Rev. Lett.}\ }\textbf {\bibinfo {volume} {102}},\ \bibinfo {pages} {064501}
  (\bibinfo {year} {2009})}\BibitemShut {NoStop}%
\bibitem [{\citenamefont {Glazier}\ \emph {et~al.}(1999)\citenamefont
  {Glazier}, \citenamefont {Segawa}, \citenamefont {Naert},\ and\ \citenamefont
  {Sano}}]{glazier1999}%
  \BibitemOpen
  \bibfield  {author} {\bibinfo {author} {\bibfnamefont {J.~A.}\ \bibnamefont
  {Glazier}}, \bibinfo {author} {\bibfnamefont {T.}~\bibnamefont {Segawa}},
  \bibinfo {author} {\bibfnamefont {A.}~\bibnamefont {Naert}}, \ and\ \bibinfo
  {author} {\bibfnamefont {M.}~\bibnamefont {Sano}},\ }\href@noop {} {\bibfield
   {journal} {\bibinfo  {journal} {Nature}\ }\textbf {\bibinfo {volume}
  {398}},\ \bibinfo {pages} {307} (\bibinfo {year} {1999})}\BibitemShut
  {NoStop}%
\bibitem [{\citenamefont {Sel'Kov}(1968)}]{sel1968}%
  \BibitemOpen
  \bibfield  {author} {\bibinfo {author} {\bibfnamefont {E.~E.}\ \bibnamefont
  {Sel'Kov}},\ }\href@noop {} {\bibfield  {journal} {\bibinfo  {journal} {Eur.
  J. Biochem.}\ }\textbf {\bibinfo {volume} {4}},\ \bibinfo {pages} {79}
  (\bibinfo {year} {1968})}\BibitemShut {NoStop}%
\bibitem [{\citenamefont {Strogatz}(2018)}]{strogatz2018}%
  \BibitemOpen
  \bibfield  {author} {\bibinfo {author} {\bibfnamefont {S.~H.}\ \bibnamefont
  {Strogatz}},\ }\href@noop {} {\emph {\bibinfo {title} {Nonlinear dynamics and
  chaos: With applications to physics, biology, chemistry, and engineering}}}\
  (\bibinfo  {publisher} {CRC press},\ \bibinfo {year} {2018})\BibitemShut
  {NoStop}%
\bibitem [{\citenamefont {Landau}\ and\ \citenamefont
  {Lifshitz}(2013)}]{landau-fluid}%
  \BibitemOpen
  \bibfield  {author} {\bibinfo {author} {\bibfnamefont {L.~D.}\ \bibnamefont
  {Landau}}\ and\ \bibinfo {author} {\bibfnamefont {E.~M.}\ \bibnamefont
  {Lifshitz}},\ }\href@noop {} {\emph {\bibinfo {title} {Fluid Mechanics}}}\
  (\bibinfo  {publisher} {Elsevier},\ \bibinfo {year} {2013})\BibitemShut
  {NoStop}%
\bibitem [{\citenamefont {Ciliberto}\ and\ \citenamefont
  {Gollub}(1984)}]{gollub1984}%
  \BibitemOpen
  \bibfield  {author} {\bibinfo {author} {\bibfnamefont {S.}~\bibnamefont
  {Ciliberto}}\ and\ \bibinfo {author} {\bibfnamefont {J.~P.}\ \bibnamefont
  {Gollub}},\ }\href@noop {} {\bibfield  {journal} {\bibinfo  {journal} {Phys.
  Rev. Lett.}\ }\textbf {\bibinfo {volume} {52}},\ \bibinfo {pages} {922}
  (\bibinfo {year} {1984})}\BibitemShut {NoStop}%
\bibitem [{Note2()}]{Note2}%
  \BibitemOpen
  \bibinfo {note} {This was suggested by one of the anonymous
  reviewers.}\BibitemShut {Stop}%
\bibitem [{\citenamefont {Waleffe}\ \emph {et~al.}(2015)\citenamefont
  {Waleffe}, \citenamefont {Boonkasame},\ and\ \citenamefont
  {Smith}}]{waleffe2015}%
  \BibitemOpen
  \bibfield  {author} {\bibinfo {author} {\bibfnamefont {F.}~\bibnamefont
  {Waleffe}}, \bibinfo {author} {\bibfnamefont {A.}~\bibnamefont {Boonkasame}},
  \ and\ \bibinfo {author} {\bibfnamefont {L.~M.}\ \bibnamefont {Smith}},\
  }\href@noop {} {\bibfield  {journal} {\bibinfo  {journal} {Phys. Fluids}\
  }\textbf {\bibinfo {volume} {27}},\ \bibinfo {pages} {051702} (\bibinfo
  {year} {2015})}\BibitemShut {NoStop}%
\bibitem [{\citenamefont {Maykut}\ and\ \citenamefont
  {Untersteiner}(1971)}]{MU71}%
  \BibitemOpen
  \bibfield  {author} {\bibinfo {author} {\bibfnamefont {G.~A.}\ \bibnamefont
  {Maykut}}\ and\ \bibinfo {author} {\bibfnamefont {N.}~\bibnamefont
  {Untersteiner}},\ }\href@noop {} {\bibfield  {journal} {\bibinfo  {journal}
  {J. Geophys. Res.}\ }\textbf {\bibinfo {volume} {76}},\ \bibinfo {pages}
  {1550 } (\bibinfo {year} {1971})}\BibitemShut {NoStop}%
\end{thebibliography}%

\end{document}